\documentclass[12pt]{iopart}
\usepackage{iopams}
\usepackage{graphicx}
\begin{document}
\newcommand{\CD}{{\cal D}}
\newcommand{\CF}{{\cal F}}
\newcommand{\CE}{{\cal E}}
\newcommand{\CR}{{\cal R}}
\newcommand{\CH}{{\cal H}}
\newcommand{\CM}{{\cal M}}
\newcommand{\CC}{{\cal C}}
\newcommand{\CQ}{{\cal Q}}
\newcommand{\average}[1]{\langle #1 \rangle_\CD}
\newcommand{\averageM}[1]{\langle #1 \rangle_\CM}
\newcommand{\averageE}[1]{\langle #1 \rangle_\CE}
\newcommand{\laverage}[1]{\langle #1 \rangle_{\CD_{\rm \bf i}}}
\newcommand{\gaverage}[1]{\langle #1 \rangle_{\Sigma}}
\newcommand{\initial}[1]{{#1_{\rm \bf i}}}
\newcommand{\now}[1]{{#1_{\rm \bf 0}}}
\newcommand{\inI}{{I}}
\newcommand{\inII}{{II}}
\newcommand{\inIII}{{III}}

%\begin{titlepage}
\title[Curvature of the present--day Universe]{On the curvature of the 
present--day Universe}

\author{Thomas Buchert$^1$ and Mauro Carfora$^{2,3}$}

\address{$^1$Universit\'e Lyon~1, Centre de Recherche Astrophysique de Lyon, CNRS UMR 5574,\\ \quad
9 avenue Charles Andr\'e, F--69230 Saint--Genis--Laval, France}

\address{$^2$ Dipartimento di Fisica Nucleare e Teorica,
Universit\`{a} degli Studi di Pavia, \\ \quad
via A. Bassi 6, I--27100 Pavia, Italy \\
$^3$Istituto Nazionale di Fisica Nucleare, Sezione di Pavia, \\ \quad
via A. Bassi 6, I--27100 Pavia, Italy\\\vspace{3pt}
Emails: buchert@obs.univ--lyon1.fr and mauro.carfora@pv.infn.it}

\begin{abstract}
We discuss the effect of curvature and matter inhomogeneities on the averaged scalar curvature 
of the present--day Universe. Motivated by studies of averaged inhomogeneous cosmologies, we 
contemplate on the question whether it is sensible to assume that curvature averages out
on some scale of homogeneity, as implied by the standard concordance model of cosmology, or
whether the averaged scalar curvature can be largely negative today, as required for an explanation 
of Dark Energy from inhomogeneities. 
We confront both conjectures with a detailed analysis of the kinematical backreaction term and 
estimate its strength for a multi--scale inhomogeneous matter and curvature distribution. Our main 
result is a formula for the spatially averaged scalar curvature involving quantities that are all 
measurable on regional (i.e. up to 100 Mpc) scales. 
We propose strategies to quantitatively evaluate the formula, and pinpoint the assumptions implied 
by the conjecture of a small or zero averaged curvature. We reach the conclusion that the standard 
concordance model needs fine--tuning in the sense of an assumed equipartition law for curvature 
in order to reconcile it with the estimated properties of the averaged physical space, whereas a 
negative averaged curvature is favoured, independent of the prior on the value of the cosmological 
constant.

\end{abstract}
% 04.20.-q  Classical general relativity
% 04.20.Cv Fundamental problems and general formalism in general relativity
% 04.40.-b   Self-gravitating systems; continuous media and classical fields in
%                   curved spacetime
% 95.30.-k   Fundamental aspects of astrophysics
% 95.36.+x  Dark energy
% 98.80.Es  Observational cosmology including Hubble constant, distance scale,
%                   cosmological constant, early Universe, etc
% 98.80.Jk  Mathematical and relativistic aspects of cosmology
%

%\end{titlepage}

\section{The debate on the averaging problem and Dark Energy}

Homogeneous and isotropic solutions of Einstein's laws of gravitation do not account for inhomogeneities in the
Universe. 
The question whether they do {\it on average} is a longstanding issue 
that is the subject of considerable debate especially in the recent literature (\cite{rasanen:darkenergy}, \cite{kolbetal} 
and follow--up references; comprehensive
lists may be found in the reviews \cite{ellisbuchert}, \cite{rasanenrev} and \cite{buchert:review}).
Averaging the scalar parts of Einstein's equations on space--like hypersurfaces of a foliation of spacetime
\cite{buchert:grgdust,buchert:grgfluid,buchert:static} 
it was found that the Friedmannian
framework is still applicable, however, one must include additional source terms due to the 
backreaction of inhomogeneities on a homogeneous--isotropic solution. These terms have
geometrical origin and, as has been recently shown, can be represented by a minimally
coupled scalar field component, a so--called {\it morphon field} \cite{morphon}, if those geometrical terms are interpreted as effective
sources in a cosmological model with Friedmannian kinematics. 
This effective field can, like quintessence, other scalar field models \cite{copeland:darkenergy}, e.g. models motivated
by higher--order Ricci curvature Lagrangians \cite{riccilagrangians}, \cite{capo}
or string--motivated effective actions \cite{string:review},  be employed to model Dark Energy.

\noindent While the Newtonian and post--Newtonian frameworks suppress these effective scalar field degrees
of freedom by construction \cite{buchertehlers}, \cite{wald}, and so cannot lead to an
explanation of Dark Energy, general relativity not only offers a wider range
of possible cosmologies, since it is not constrained by the assumption of Euclidean or
constant curvature geometry and small deviations thereof, but it is also needed to describe an effect that is strictly absent in
a Newtonian model and in a standard (quasi--Newtonian) perturbation approach at a fixed background. This effect is reflected by the 
coupling of the fluctuations to the averaged model. In other words, fluctuations may be small, but measured relative to a non--Friedmannian
background, and the evolution of this latter is most clearly expressed in terms of the evolution of effective geometrical properties 
such as the averaged scalar curvature we are considering in this paper  (see \cite{buchert:review} for detailed explanations). 
Speaking in favour of an averaged cosmology, it certainly enjoys the more
physical status of incorporating inhomogeneities, and the clearcut fact that the effect of these
inhomogeneities can be modelled by a scalar field speaks, by William of Ockham's {\em razor}, 
against introducing a dominating cosmological constant or an extra fundamental scalar field that is known 
to violate energy conditions in order to explain observational data. 
From this point of view one would also conclude that perturbation theory, if formulated at the background 
of a FLRW model with a dominating cosmological constant or an external scalar field source, 
would also not account for the physics behind the Dark Energy component.  

On the other hand, the FLRW cosmology provides a remarkably successful {\em  fitting model}
to a large number of observational data. As already mentioned, 
the price to pay is an unclear physical origin of either a dominating
cosmological constant or an extra scalar field source that dominates recently.
Given the fact that also a large amount of sources in the form of Dark Matter has not yet been detected in 
(non--gravitational) experiments, 
the standard model parametrizes an overwhelming fraction (95 percent) of physical ignorance. 
The generally held view, however, is that the FLRW cosmology indeed describes the {\em physical
Universe} on average, which -- if true -- in turn asks for either a modification of the laws of gravitation, or 
the postulation of the above--mentioned dark sources of yet unknown origin. 
Moreover, the widespread use of the wording `fitting model' is just name--dropping
unless we devise a way to explicitly construct a smooth metric out of the inhomogeneous distributions of matter
and curvature \cite{ellis,ellisstoeger}. In this -- more refined -- sense the FLRW cosmology is not a fitting model, 
rather {\em it furnishes a conjecture on integral properties of the physical Universe} that, as we believe, 
has to be first verified or falsified before more exotic vehicles of explanation are invoked.

Both, the FLRW cosmology and a backreaction--driven averaged cosmology are candidates for the
description of these integral properties, and in this paper we shall estimate
these properties from regionally (up to, say, 100 Mpc) observable quantities.
For the FLRW cosmology in the form of the {\em concordance model} \cite{lahav}, \cite{SNLS}, \cite{spergeletal,komatsuetal}, the physical
model is described by on average vanishing scalar curvature, while for a backreaction--driven cosmology,
if we expect that Dark Energy can be fully routed back to inhomogeneities, the issue appears open. However,
the consequences of a backreaction--driven model have been qualitatively, and to the extent we need also
quantitatively, exploited in a number of recent papers (see \cite{buchert:review} and references therein). 
For example, since a quantitative estimation of kinematical backreaction depends on specifying an evolution
model for the inhomogeneities, the analysis of exact solutions like the 
Lema\^\i tre--Tolman--Bondi solution (see, e.g., \cite{celerier1}, \cite{singh1}, \cite{LTBluminosity2}, \cite{teppo07}, and the reference lists in \cite{rasanenrev} and \cite{buchert:review}), 
or scaling laws that satisfy the averaged equations \cite{morphon} have been investigated. 
{\em Consistency} of an explanation of Dark Energy with the framework of the averaged equations has been demonstrated
for both globally homogeneous cosmologies \cite{morphon} (an assumption that we also adopt in the present
paper) and, alternatively,   
globally inhomogeneous cosmologies \cite{buchert:darkenergy,buchert:static}. 
Although a quantitative evaluation of the backreaction effect in a generic inhomogeneous model is still to come, we 
already know a few features of a `working' model, which are enough for our considerations.

\begin{figure}[h]
\begin{center}
\includegraphics[bb= 0 0 540 470,scale=.5]{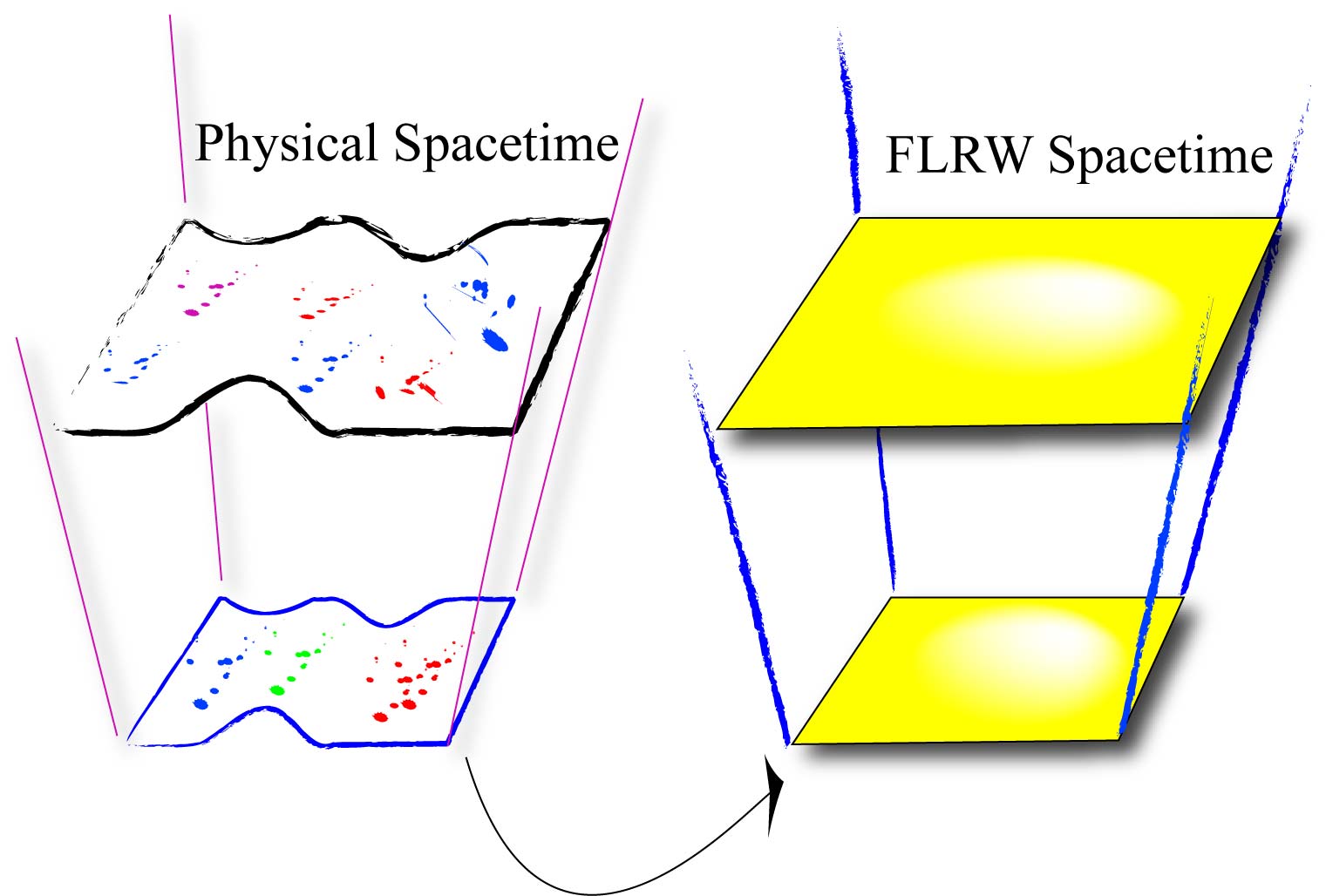}
\caption{FLRW cosmology provides a successful fitting to observational data. The price to pay is an unclear origin of either a dominant cosmological constant or of an extra scalar field source that appears to dominate the present dynamics of the Universe.
Inhomogeneous hypersurfaces may be subjected to a smoothing procedure in order to find
the corresponding smooth, i.e. constant--curvature fitting model that we may call a 
{\em FLRW template}. We would then consider a hypersurface at a given instant of time;
we cannot expect that the time--evolved inhomogeneous model can be mapped to a 
the constant--curvature evolution of the FLRW cosmology.
}
\end{center}
\end{figure}

We do not aim at investigating a `fitting model' (see Figure 1) for the present--day Universe
in the strict sense mentioned above; this is the subject of ongoing work. 
We only remark here that any model for the evolution of the averaged variables
can be subjected to a smoothing procedure in order to find
the corresponding smooth, i.e. constant--curvature fitting model that we may call a 
{\em FLRW template}. We would then consider a hypersurface at a given instant of time;
we cannot expect that the time--evolved inhomogeneous model can be mapped to a 
constant--curvature model.
The above two candidates would provide different starting points, i.e. the initial data of a smoothing procedure for a given hypersurface are different.
It was recently argued \cite{singh2} 
that the averaged universe model (now both kinematically and
geometrically averaged) could be represented by an effective 
FLRW metric with a time--scaling factor that differs from the usual global scale 
factor of a homogeneous--isotropic model and which is 
determined by the kinematically averaged Einstein equations.
This ansatz for an effective metric assumes that smoothing (i.e. spatial rescaling)
the actual matter and curvature inhomogeneities
does not leave traces in the smoothed--out FLRW template metric {\em at all times}. 
In a forthcoming paper we are going to analyze this assumption in detail employing previous results on an
explicit smoothing algorithm \cite{hamilton:ricciflow2,carfora:deformation1,carfora:RG,klingon,buchertcarforaPRL,buchertcarfora:quartet,
carforabuchert:perelman}.
We emphasize that we are entitled to investigate integral properties of physical variables on a given hypersurface {\em without} 
entering the different question
of whether this hypersurface (if actively deformed) can be effectively described by a `best--fit' constant--curvature geometry.

A rough guide that helps to understand the motivation of the present work is the following.
For small inhomogeneities in the matter and curvature distributions,  an approximate description
of the cosmic evolution {\em on average} by a homogeneous solution of Einstein's
laws of gravitation may be fine,
but for the Late Universe featuring strong inhomogeneities, the validity of this approximation is not evident. 
We are going to address and 
justify this remark in the present work {\em on the assumptions} that (i) a homogeneous
model satisfactorily describes the early stages of the matter--dominated epoch and (ii) there exists a scale of homogeneity.

\bigskip

We proceed as follows. In Section~\ref{section:phenomenology} we look at the present--day
Universe and device a three--scale model for it, where we hope that both readers, those who
advocate the standard picture of the concordance model and those who advocate a
backreaction--driven cosmology, agree. Then, in Section~\ref{section:multiscaling},
we implement the details of this multi--scale picture and reduce the determining sources to those that
are in principle measurable on regional (i.e. up to 100 Mpc) scales. Detailed estimates of the kinematical
backreaction and averaged scalar curvature  follow in Section~\ref{section:Q}, where 
we also provide simplified estimates in the form of robust bounds by, e.g., restricting the measurement of fluctuations to a comoving frame.
In Section~\ref{section:conclusions} we confront this latter result with the different
assumptions on the actual averaged scalar curvature of the present--day Universe.

\section{A fresh look at the present--day Universe}
\label{section:phenomenology}

\subsection{Phenomenology: the volume--dominance of underdense regions}

Observing the Universe at low redshift returns the impression of large volumes that are
almost devoid of any matter; a network of large--scale structure surrounds these {\em voids}
that seem to be hierarchically nested and their sizes, depending on their definition, range
from regional voids with less than 10 percent galaxy number content of the order of ten Megaparsecs 
\cite{hoyle}, \cite{furlanetto}, to relatively thinned--out regions of larger number density that,
if smoothed, would span considerable volume fractions of currently available large--scale
structure surveys. While the overall volume of the observable Universe seems to be dominated by
underdense regions (the particular value of the volume--fraction of underdense regions being dependent on the threshold of this underdensity), the
small fraction of the volume hosting overdensities (groups, clusters and superclusters of galaxies) is
itself sparsely populated by luminous matter and, this latter, appears as a highly nonlinear `spiky' distribution. 
The phenomenological impression that
matter apparently occupies a tiny fraction of space {\em at all length scales} could be
questioned by saying that {\em Dark Matter} might be more smoothly distributed. Also,
clusters contain a large amount of intergalactic gas, there are non--shining baryons, etc.,
so that the notion of an `underdense
region' has to be treated with care. However, simulations of {\em Cold Dark Matter}, 
assumed to rule the
formation of large--scale structure, also demonstrate that voids dominate the 
present--day distribution. Again depending on the particular definition of a void, their fraction of
volume occupation could, to give a rough value, be conservatively quoted as being 60 \cite{colberg}  
percent in standard {\em $\Lambda-$Cold Dark Matter} simulations counting strong underdensities, and is certainly larger for 
more densly populated but still underdense regions.

Thinking in terms of a {\em homogeneous model} of the Universe (not necessarily a homogeneous
{\em solution}), i.e. a distribution of
matter that on average does not depend on scale beyond a certain large scale (the scale
of homogeneity), one would paint the picture of a redistribution of matter due to nonlinear
gravitational instability. In a Newtonian simulation (where an eventually constant curvature
of a FLRW spacetime is factored out on a periodic scale)
this would happen in such a way that, due to the preservation
of the overall material mass, an {\em equipartition} of overdense small--volume regions and
underdense large--volume regions with respect to the mass content results, so that a
sensible spatial average of the matter distribution must comply with the original value
of the homogeneous density. In other words, the assumption that a volume--averaged distribution of matter would be compatible with a homogeneous model
of the same average density seems to be a robust assumption, especially if inhomogeneities are dynamically generated out of an almost homogeneous
distribution. This picture is true in Newtonian simulations, but
for a subtle reason: although
the time--evolution of the averaged density as a result of {\em non--commutativity}
of evolution (time--scaling) and spatial averaging gives rise to kinematical backreaction,
the periodic architecture and the Euclidean geometry of a Newtonian cosmology  
imply that these additional terms have to vanish (see \cite{buchertehlers} for a detailed discussion of all these issues and proofs).
In a general--relativistic framework this picture is in general false, even at one instant of time: 
the reason is that a Riemannian volume average incorporates the 
volume measure that is different for negatively and positively curved domains of
averaging, and curvature does not obey an equipartition law (see Figure 2).

Note also that a volume averaging on a Riemannian 3--surface could, even on the largest scale, 
introduce a {\em volume effect} in the comparison of the volume of a constant--curvature space and
the actual volume of an inhomogeneous hypersurface
(see \cite{buchertcarforaPRL} and \cite{buchertcarfora:quartet} for the definition and discussion
of the {\em volume effect}; see also Hellaby's volume--matching example \cite{hellaby:volumematching}). 
The standard model (but also the recent suggestion by \cite{singh2}) implies that there is no
such effect on large scales. It is illustrative to think of 2--surfaces, where curvature inhomogeneities always add up in the
calculation of the  total surface, so that there certainly is a large {\em 2--volume effect} due to surface roughening, 
but for three--dimensional manifolds,
negative and positive curvature contribute with opposite signs, and so the {\em 3--volume effect} cannot easily be quantified. 

\begin{figure}[h]
\begin{center}
\includegraphics[bb= 0 0 540 470,scale=.4]{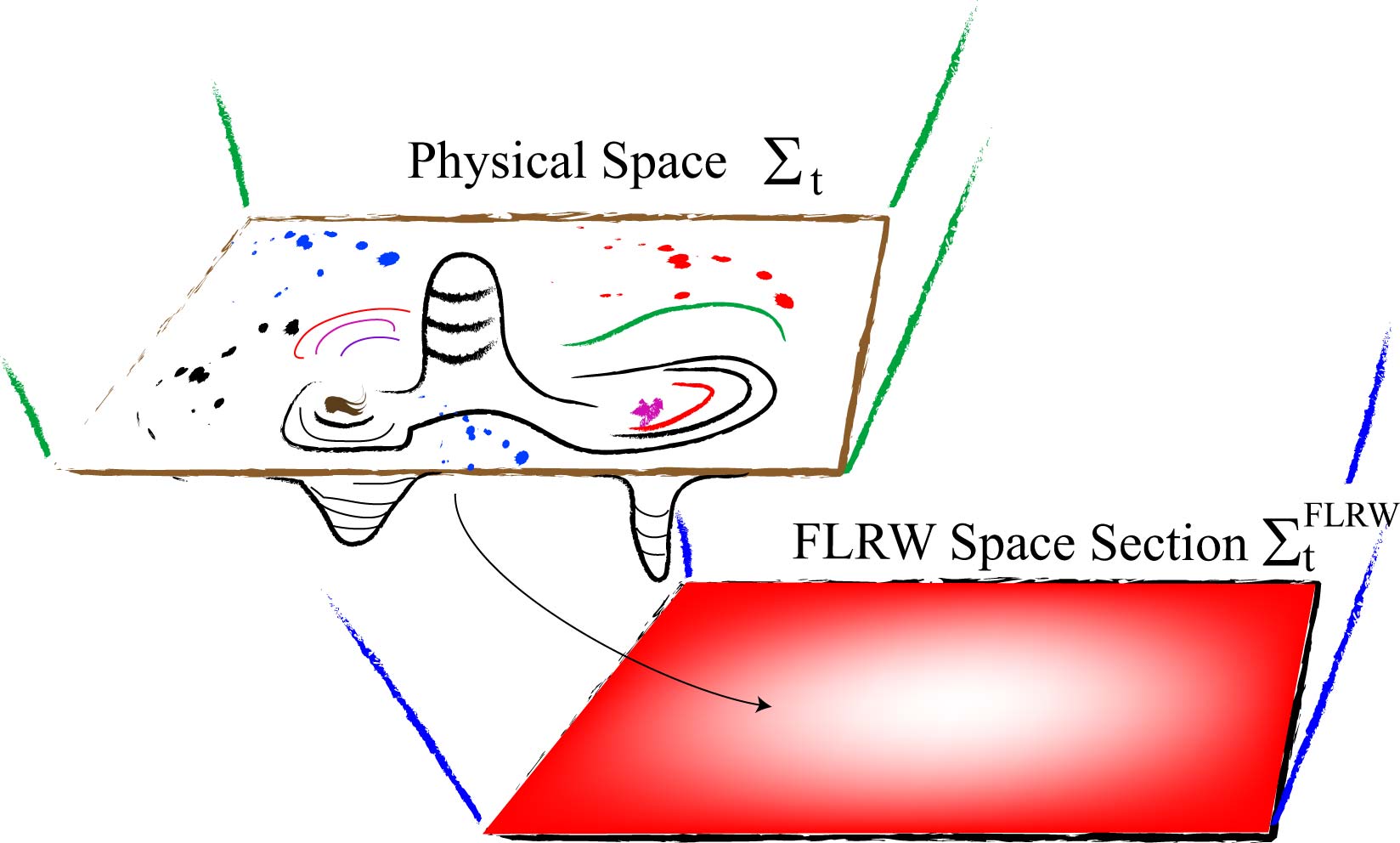}
\caption{The picture that the scalar curvature in the physical space would average out on some large scale of homogeneity is naive in a number of ways.
There is no equipartition law for the scalar curvature that would be dynamically preserved.}
\end{center}
\end{figure}

Given the above remarks, one is no longer tempted to draw a picture of equipartition 
for the intrinsic curvature distribution.
We shall, in this paper, not discuss the time--evolution of the scalar curvature (see \cite{rasanenrev} and \cite{buchert:review} for detailed
illustrations and discussions), large--time asymptotics \cite{reiris1,reiris2}, the role of a constant--curvature parameter 
in the fit to observations \cite{bruceSN}, or curvature models \cite{boundsoncurvature,rasanenk(t)} (that are all related subjects of interest),
but instead contemplate on the distribution of curvature at one given instant of time. Here, we
demonstrate that the picture we would wish to establish in the concordance model, 
namely that the scalar curvature would average out on some large scale of homogeneity, 
is naive in a number of ways, and we shall implement the geometrical
aspects of such a picture in Section~\ref{section:multiscaling}.
Obviously, this issue cannot be addressed with Newtonian simulations; 
the curvature degree of freedom is
simply absent. We know that in Riemannian geometry negatively curved regions 
have a volume that is larger than the corresponding
volume in a Euclidean space section, and positively curved regions have a 
smaller volume, thus enhancing the actual volume fraction of underdense regions.

We are now going to develop a multi--scale picture of the present--day Universe that 
is useful in the context of quantitative estimates, and that also helps to quantify multi--scale dynamical models,
e.g. the one proposed by Wiltshire and his collaborators 
\cite{wiltshire05a,wiltshire05b,wiltshire07a,wiltshire07b,wiltshire07c,wiltshire07d},
but it relates as well to any model that involves considerations of structures on different spatial scales.

\subsection{Scaling and coarse--graining: multi--scale picture of the Universe}

We are going to introduce three spatial scales. 
First, a {\em scale of homogeneity} $L_\CH$ that could be identified with the size of
a compact universe model, but need at least be larger than the largest observed typical structures.
Second, a scale $L_\CE$ that is as large
as a typical void (within a range of values that depends on our definition of a devoid region), 
and third, a scale $L_\CM$ that is large enough to host typical 
bound large--scale objects such as a rich cluster of galaxies. 
In observational cosmology we strictly have 
\begin{equation}
L_\CH\;>\;L_\CE\;>\;L_\CM\;.
\end{equation}
For the first we may also think of a length of the order of the Hubble--scale. 
The lower bound on this
scale not only depends on the statistical measure with respect to which one considers
the matter and curvature distributions as being {\em homogeneous}, but also on the concept of {\em homogeneity}
that we have in mind. We here imply that averages {\em of any variable}
beyond this scale will in practice
no longer depend on scale, while generically, this may not happen at all. 
We do not claim here that this is indeed true, but we adopt
this point of view in order to have a more transparent way of comparison with the standard
model of cosmology. 
The assumption of existence of a
scale of homogeneity may be a strong hypothesis; it is our choice of restricting the 
generality of the problem.

\begin{figure}[h]
\begin{center}
\includegraphics[width=\textwidth]{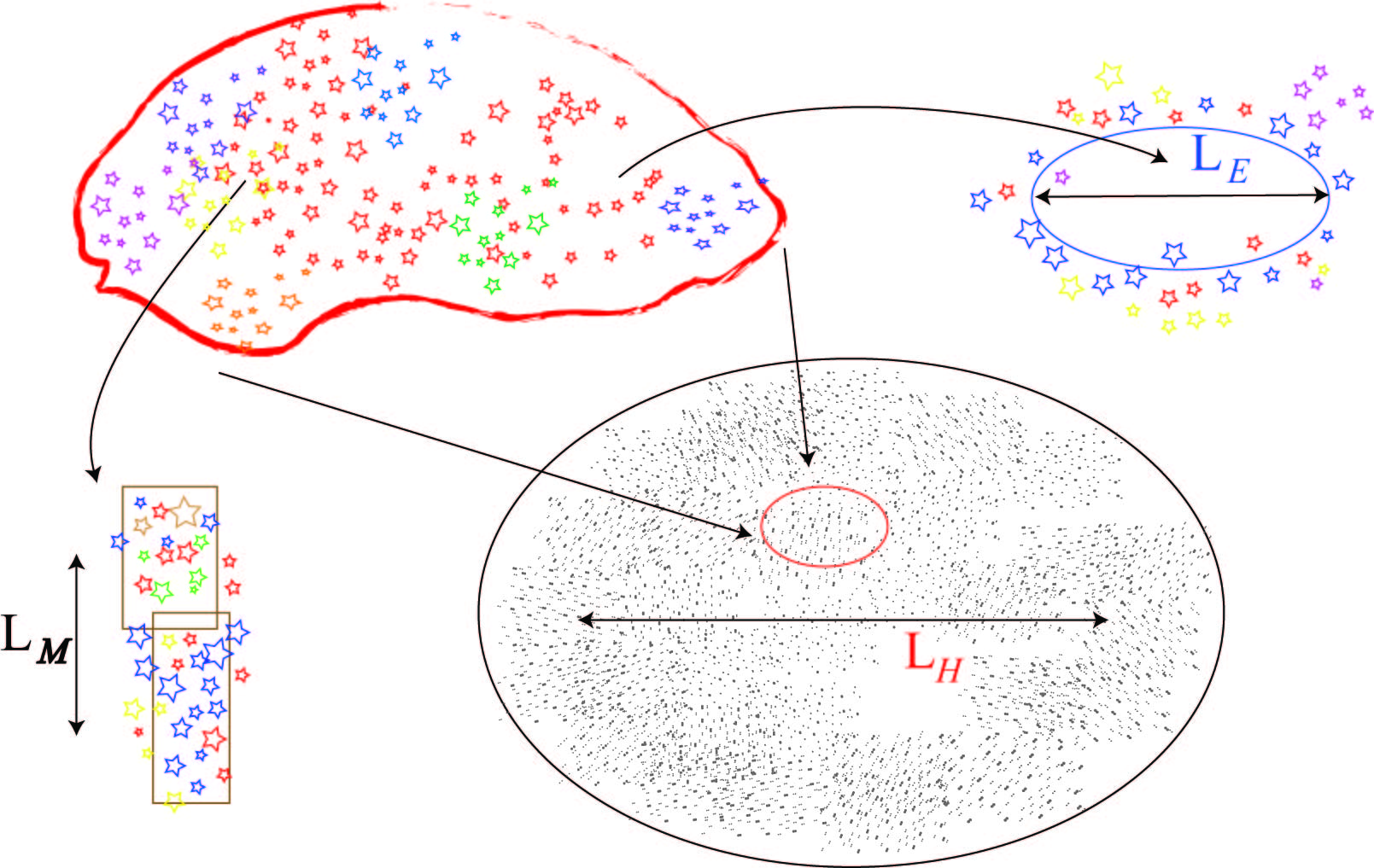}
\caption{The three spatial scales to be used in our multi--scale picture of the
Universe: the scale of homogeneity $L_{\mathcal{H}}$, the scale of a typical void $L_{\mathcal{E}}$, and
the scale $L_{\mathcal{M}}$ hosting typical bound large--scale objects, \emph{e.g.} a rich cluster of galaxies.}
\end{center}
\end{figure}

According to what has been said above, we are entitled to assign different properties to the different scales,
and we shall also sometimes 
idealize these properties in order to construct a simple but flexible model that reflects the
phenomenology described above in terms of a small set of parameters.

We start with an overview of the basic ingredients of our model and postpone details
to Section~\ref{section:multiscaling}.
We employ the Hamiltonian constraint (see (\ref{constraints}) below), 
spatially averaged on a given domain ${\cal D}$ (Scale $L_\CH$) that covers 
a union of underdense regions ${\cal E}$ (Scale $L_\CE$) and occupied overdense 
regions ${\cal M}$ (Scale $L_\CM$).
We write the Hamiltonian constraint averaged over the first scale 
(for details see below and, e.g. \cite{buchert:static}):
\begin{equation}
\label{averagehamiltonD}
{\average{\CR}\;=\; -6H_\CD^2 -{\cal Q}_\CD} + {16\pi G}\langle\varrho\rangle_\CD
+2\Lambda\;,
\end{equation}
with the  total restmass   $M_\CD :=\langle\varrho\rangle_\CD |{\cal D}|$.
The  averaged   spatial scalar curvature is denoted by 
$\average{\CR}$, $H_\CD :=1/3 \langle\theta\rangle_\CD$ abbreviates the averaged rate of expansion $\theta$ in terms of 
a {\em volume Hubble rate},  and   
the  {\it kinematical backreaction  term}  ${\cal Q}_\CD$ encodes inhomogeneities in the extrinsic curvature distribution
(or the kinematical variables); it is detailed in 
Section~\ref{section:multiscaling} (see Figures 3 and 4).

\begin{figure}[h]
\begin{center}
\includegraphics[width=\textwidth]{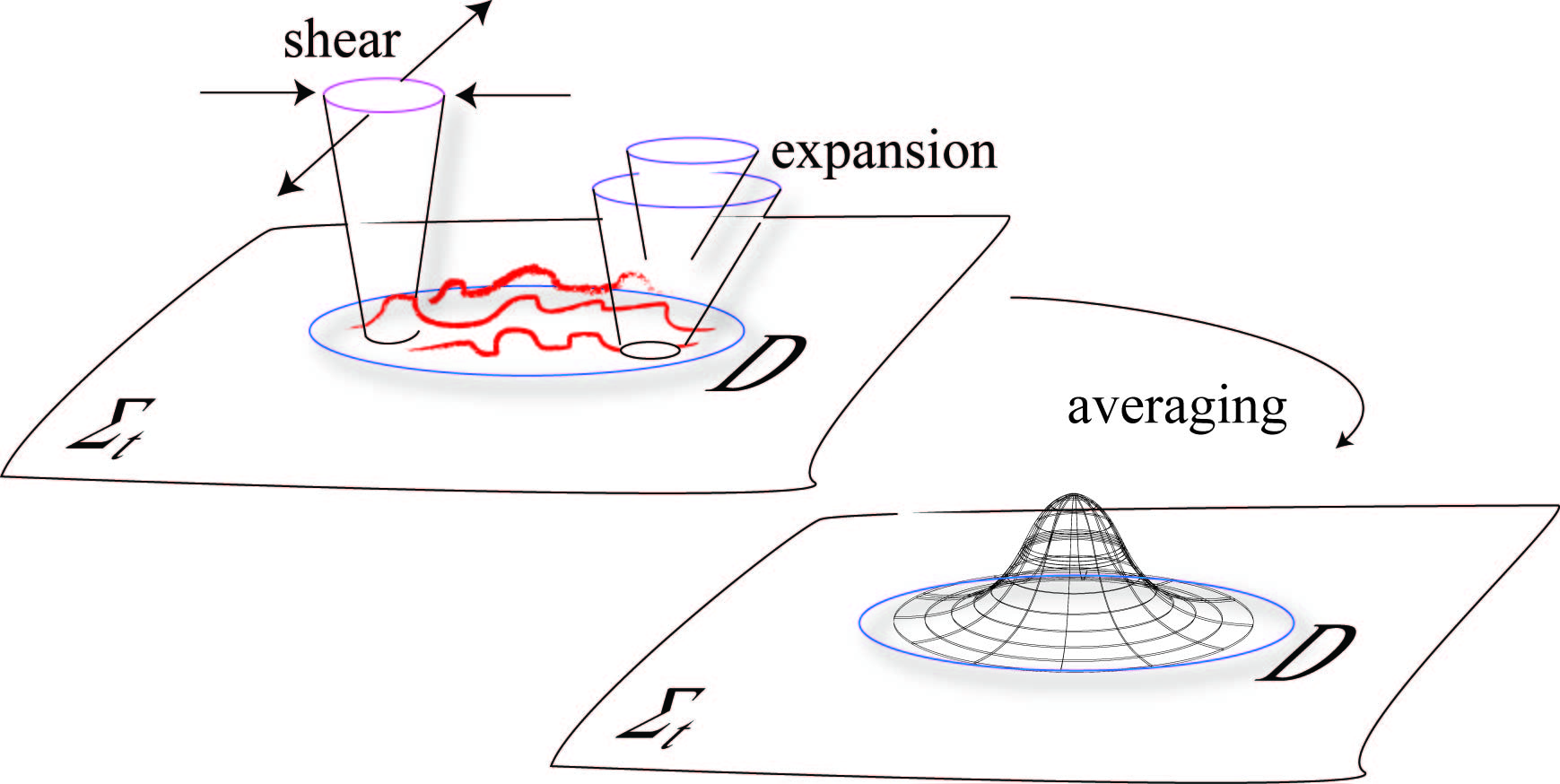}
\caption{The  fluctuations in the local expansion and shear generate a kinematical backreaction.
This acts as a source term for scalar curvature when averaging the Hamiltonian constraint over the region of near--homogeneity $\mathcal{D}$.}
\end{center}
\end{figure}

Now, let us consider for illustrative purposes an idealization on the two other scales (in our concrete calculations later
we shall indicate clearly when we make use of it): we require the 
volume Hubble expansion to be subdominant in matter--dominated regions and,
on the other hand,  the averaged density to be subdominant in devoid regions.
In the first case, an expansion or contraction would contribute negatively to the averaged scalar curvature and so would, e.g., 
enhance a negative
averaged curvature; in the second case, the presence of a low averaged density would contribute positively.
We can therefore reasonably expect that, whether we use a strong idealization (see below) or a weaker distinction between over-- and underdense regions, the
overall argument based on the existence of such a partitioning enjoys some robustness.  
We shall also be able to condense our assumption on the partition between over-- and underdense regions by introducing a parameter for the occupied volume fraction,
$\lambda_\CM := |\CD_\CM| / |\CD|$, where $|\CD_\CM|$ denotes the total volume of the union
of occupied regions $\CM$; its value
may be chosen more conservatively to weaken an eventually unrealistic idealization. At any rate, we shall keep our calculations as general as possible 
before we eventually invoke an idealization for illustrative purposes; it is only this latter quantity $\lambda_\CM := |\CD_\CM| / |\CD|$ that parametrizes the
(geometrical) volume--partitioning of the distributions of inhomogeneities.

Note that, if we would strictly idealize voids to have $\langle\varrho\rangle_\CE = 0$ and
and matter--dominated regions to have $H_\CM = 0$ for $\lambda_\CM<\lambda^{cr}_\CM$, where $\lambda^{cr}_\CM \ll 1$ is some critical scale (see 
Section~\ref{section:multiscaling} for more details on this transitory scale and the controling of this idealization),
then we would have:
\begin{equation}
\label{averagehamiltonE}
\averageE{\CR}\;=\;-6H_\CE^2  -{\cal Q}_\CE +2\Lambda\;,
\end{equation}
and
\begin{equation}
\label{averagehamiltonM}
{\averageM{\CR}\;=\;-{\cal Q}_\CM} + {16\pi G}\langle\varrho\rangle_\CM +2\Lambda\;,
\end{equation}
together with
\begin{equation}
H_\CD = (1-\lambda_\CM ) H_\CE \qquad{\rm and}\qquad \langle\varrho\rangle_\CD =
\lambda_\CM \langle\varrho\rangle_\CM\;\;.
\label{relationaverages}
\end{equation}
This simplified view is useful as a rough guide on the sign of the averaged scalar curvature: consider for example the case where
the kinematical backreaction terms in the above 
equations are quantitatively negligible, and let us put $\Lambda =0$; we then infer that the averaged scalar curvature must be {\em negative} on 
Scale $L_\CE$ and {\em positive} on Scale $L_\CM$, what obviously complies with what we expect.
A non--vanishing $\Lambda > 0$ is employed in the concordance model to compensate the negative curvature.

\subsection{Cosmological parameters and scalar curvature}

For our discussion we introduce a set of adimensional average characteristics 
that we define for the largest scale:
\begin{equation}
\label{omega}
\fl\qquad
\Omega_m^{\CD} : = \frac{8\pi G}{3 H_{\CD}^2} \langle\varrho\rangle_{\cal D}  \;\;;\;\;
\Omega^\CD_{\Lambda}:= \frac{\Lambda}{3H_\CD^2}\;\;;\;\;
\Omega_{\cal R}^{\CD} := - \frac{\average{\cal R}}{6 H_{\CD}^2 }\;\;;\;\;
\Omega_{\cal Q}^{\CD} := - \frac{{\cal Q}_\CD}{6 H_{\CD}^2 } \;\;.
\end{equation}
We shall, henceforth, call these characteristics `parameters', but the reader should keep in 
mind that indexed variables are scale--dependent functionals.
Expressed through these parameters the averaged Hamiltonian constraint 
(\ref{averagehamiltonD}) on the scale $L_\CH$ assumes the form:
\begin{equation}
\label{hamiltonomegaD}
\Omega_m^{\CD}\;+\;\Omega^\CD_{\Lambda}\;+\;\Omega_{\cal R}^{\CD}\;+\;
\Omega_{\cal Q}^{\CD}\;=\;1\;\;.
\end{equation}
In this set, the averaged scalar curvature parameter and the kinematical backreaction parameter
are directly expressed through $\average{\cal R}$ and ${\cal Q}_{\CD}$, respectively.
In order to compare this pair of parameters with the `Friedmannian constant--curvature parameter' 
that is the only curvature contribution in the standard model, 
we can alternatively introduce the pair (see, e.g., \cite{buchert:static})
\begin{equation}
\label{omeganewton}
\Omega_{k}^{\CD} := - \frac{k_{\initial\CD}}{a_\CD^2 H_{\CD}^2 }\;\;;\;\;
\Omega_{{\cal Q}_N}^{\CD} := \frac{1}{3 a_\CD^2 H_\CD^2}
\int_{\initial{t}}^t \rmd t'\ {\cal Q}_\CD\frac{\rmd }{\rmd t'} a^2_\CD(t')\;\;,
\end{equation}
being related to the previous parameters by
$\;\Omega_{k}^{\CD} +\Omega_{{\cal Q}_N}^{\CD}\;=\; 
\Omega_{\cal R}^{\CD} + \Omega_{\cal Q}^{\CD}$.

\vspace{3pt}

For any of the smaller domains we discuss the corresponding adimensional parameters by dividing averages $\langle ..... \rangle_{\cal F}$ on the domain $\cal F$ always by 
$H^{2}_{\cal D}$ to avoid confusion.
This will also avoid the pathological and useless definition of the
cosmological parameters, e.g. on the domains $\CM$, where they are actually undefined
in a strict idealization, since $H_\CM$ is assumed to vanish.

To give an illustration for the scale--dependence,
note that, in the strictly idealized case,
$\Omega^\CD_m$ can be traced back to the average density in matter--dominated
regions, $\langle\varrho\rangle_\CD \cong \lambda_\CM \langle\varrho\rangle_\CM$,
and thus, inevitably, the density parameter constructed with an observed $\langle\varrho\rangle_\CM$ on the scale $\CM$ and divided by the global Hubble factor
cannot be extrapolated to the global parameter. For example, a value today of $\alpha$ for this parameter would, for a volume
fraction of matter--dominated regions of $\lambda_\CM =0.2$, result in 
$\Omega^\CD_m = 0.2 \alpha$, i.e. a substantially smaller value that compensates the missing matter in 
the regions $\CE$, if they are idealized to be empty.
Note in this context that a smaller mass density parameter on the global scale would also imply a smaller value
of the necessary amount of backreaction.
This can be seen by considering the {\em volume deceleration parameter},
\begin{equation}
q^\CD := \frac{1}{2}\Omega^\CD_m + 2\Omega^\CD_\CQ - \Omega^\CD_{\Lambda}\;\;,
\end{equation}
which, for $\Lambda = 0$, shows that decreasing the matter density parameter would also
decrease the necessary backreaction in order to find, e.g. {\em volume acceleration}
($q^\CD < 0$).

\vspace{10pt}

Let us now discuss one of the motivations of the present work related to 
the Dark Energy debate. To this end, we have to explain
why a substantial negative averaged scalar curvature is needed, so that Dark Energy could be partly or completely
routed back to inhomogeneities, i.e. $\Omega^\CD_{\Lambda} =0$ in the above equations,
and, say, also $\Omega^{\now\CD}_m \approx 0.25$\footnote{We specify the index $\now\CD$
as soon as we go to numerical estimates, indicating the value of the corresponding parameter
{\em today}. Note that, in the averaged models, we may weaken this
constraint, since we can allow for a scale--dependence of this parameter in contrast to the situation in the standard model;
see \cite{buchert:review} and \cite{teppo2} for related discussions.}
in conformity with the standard model of cosmology. 
First, by {\em substantial} we mean a cosmologically large
negative curvature, i.e. expressed in terms of the cosmological parameters 
�$\Omega_{\cal R}^{\now\CD} \approx 1$.
This parameter is exactly equal to $1$, i.e. the averaged curvature is of the order of minus
the square of the Hubble parameter, if there are no matter and no expansion and shear
fluctuations. This applies to a simplified model of a void \cite{rasanenrev}. Second, in order to
explain Dark Energy, the sum of the curvature and backreaction parameters has to add
up to mimic a cosmological constant parameter of the order of �
$\Omega^{\now\CD}_{\Lambda}\approx 0.7$ in the case of a full compensation of the cosmological constant today,
where we have adopted the assumption of the concordance model \cite{lahav} of 
$\Omega^\CD_k =0$.
A conservative scenario has been quantified in \cite{morphon} where it was assumed that 
the Universe at the Cosmic Microwave Background epoch is described by a weakly perturbed 
FLRW model and the amount of Early Dark Energy is negligible. The resulting scenario that
would create enough Dark Energy features a strong curvature evolution from a negligible
value to �$\Omega^{\now\CD}_{\cal R} \approx 1$,
while the backreaction parameter must evolve from a negligible value to
�$\Omega^{\now\CD}_{\cal Q} \approx -0.3$, 
i.e. it has to be dominated by expansion fluctuations on the scale $L_\CH$ today. This scenario implies
that the averaged scalar curvature must be close to the value of our simple void model,
while at early times the averaged curvature parameter was compatible with zero. 
Speaking in terms of a {\em morphon field} \cite{morphon}, where the scalar curvature
is associated with the potential of a scalar field, this scenario corresponds to the 
{\em phantom quintessence} sector (with negative kinetic energy of the scalar field), 
in which extrinsic curvature fluctuations ({\em kinematical backreaction}) 
grow slightly. We shall adopt this (present) value of
the curvature parameter in our analysis as an extreme candidate for a backreaction--driven cosmology.

Finally, we wish to emphasize that we are looking at spatial integral properties of inhomogeneous {\it models} of the Universe
at the present time. This study does not include the important questions of (i) how the present--day structure we are looking at evolved dynamically 
out of an earlier state, and (ii) how these integral properties would relate to deep observations that necessarily involve considerations of the
inhomogeneous lightcone and observable averages along the lightcone. We shall later propose strategies for the determination of these integral properties, also from observations. As a rule of thumb, observational results may be directly used in a shallow redshift interval, where the lightcone effect would be subdominant; galaxy catalogues as they are compared with the spatial distribution of fluctuations ``at the present time'' in simulations is an example).  Our main result is a formula for the spatially averaged scalar curvature involving quantities that are all 
measurable on regional (i.e. up to 100 Mpc) scales, and it is therefore accessible by a shallow redshift interval.  

\clearpage

The key--results of the following, necessarily technical multi--scale analysis are
Eqs.~(\ref{resultQ}) and (\ref{resultR}), and their discussion thereafter. In particular, disregarding the contribution of gravitational radiation of cosmological origin, the adimensional backreaction term $\Omega^\CD_\CQ$ can be written as (see (\ref{resultQ})):

\begin{eqnarray}
\fl
-\Omega^\CD_\CQ           \;=\;
\lambda _{\CM}\,\left[\left(1-8\,\frac{L_{\nabla\theta,\nabla\theta}^{2}}{L_{\delta H_{\CM}}^{2}}\right)\,\left(\frac{\delta ^{2}H_{\CM}}{H_{\cal D }^{\,2}}\right)-2\,V^{2}_{\varrho }[\CM]\,
\frac{L_{J,J}^{2}}{L_{\delta H_{\CM}}^{2}}\,\left(    
\frac{(8\pi G)^{2}\,\left\langle \varrho \right\rangle^{2} _{\cal M }\,  L_{\delta H_{\CM}}^{2}}{H_{\cal D }^{\,2}}\right) \right] \nonumber\\
\nonumber\\
\fl\qquad
+(1-\lambda _{\CM})\,
\left[\left(1-8\,\frac{L_{\nabla\theta,\nabla\theta}^{2}}{L_{\delta H_{\CE}}^{2}}\right)\,\left(\frac{\delta ^{2}H_{\CE}}{H_{\cal D }^{\,2}}\right)-2\,V^{2}_{\varrho }[\CE]\,
\frac{L_{J,J}^{2}}{L_{\delta H_{\CE}}^{2}}\,\left(    
\frac{(8\pi G)^{2}\,\left\langle \varrho \right\rangle^{2} _{\cal E }\,  L_{\delta H_{\CE}}^{2}}{H_{\cal D }^{\,2}}\right) \right] \nonumber\\
\nonumber\\
\fl\qquad
+{\lambda_\CM}(1-\lambda_\CM)\,\frac{\left(H_\CE-H_\CM\right)^{2}}{H_{{\cal D}}^{2}}
\pm  
\,\,2V^{2}_{\varrho }[\CD]\,\frac{L_{\nabla\theta,J}^{2}}{L_{\delta H_{\CD}}^{2}}\,
\left( \frac{32\pi G\,\left\langle \varrho \right\rangle _{\cal D } \,\;L_{\delta H_{\CD}}\,
\left(\delta ^{2}H_\CD\right)^{\frac{1}{2}}}{H_{\cal D }^{\,2}} \right) \nonumber\;,\\
\label{Mform}
\end{eqnarray}
\noindent together with the formula for the adimensional averaged scalar curvature term:
\begin{equation}
\label{resultR}
\Omega_\CR^\CD + \Omega^\CD_{\Lambda} = 1 - \Omega^\CD_m - \Omega^\CD_\CQ  \quad {\rm with}\quad \Omega^\CD_\CQ \quad {\rm from}\; {\rm above}\;,
\end{equation}
where  the rough meaning of the various terms is \emph{pictorially} described in Figure 5.
\begin{figure}[h]
\begin{center}
\fl
\includegraphics[bb= 0 0 540 470,scale=.5]{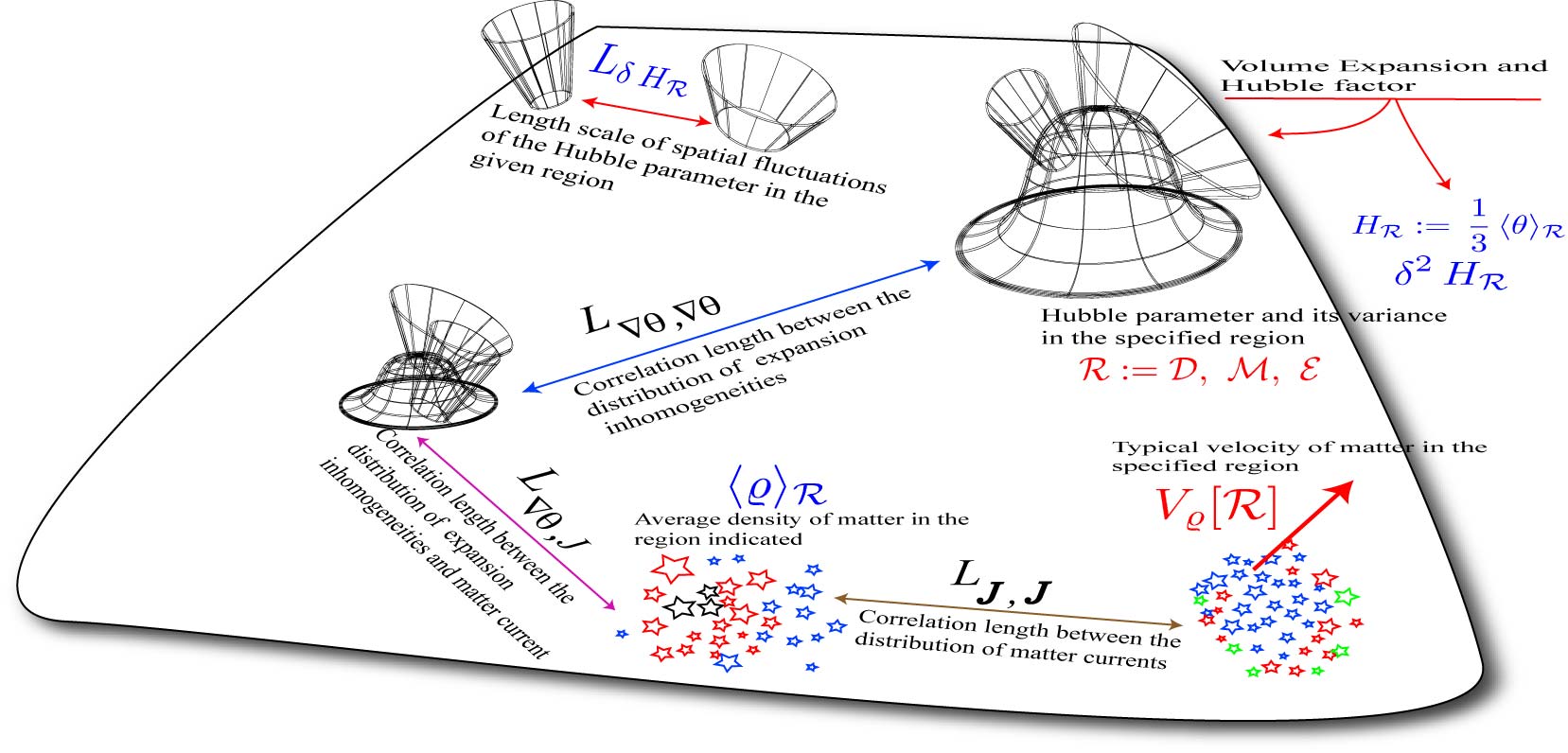}
\caption{A pictorial representation of the meaning of the various terms characterizing the expression (\ref{Mform}) for the adimensional backreaction $\Omega^\CD_\CQ$. The Hubble parameters $H_{\cal R }$ and their variances $\delta^{2}\,H_{\cal R }$, as ${\cal R}$ varies in the various regions ${\cal D}$, $\cal M$, and $\cal E$, are defined in $\S$ \ref{Hubble}. The  correlation lengths $L_{\nabla\theta,\nabla\theta}$, $L_{J,J}$, $L_{\nabla\theta,J}$ and the parameters $L_{\delta H_{\CR}}$, $V^{2}_{\varrho }[\CD]$ are characterized in $\S$ \ref{shcorrlen}.}
\end{center}
\end{figure}

\noindent
The following sections also prepare future
work, e.g. on the determination of an ``optimal frame'' in which the variables are to be averaged in an evolving cosmological hypersurface. 

\clearpage

\section{Multi--scale analysis of curvature}
\label{section:multiscaling}

\subsection{The Hamiltonian constraint}
\label{subsection:hamilton}

In order to discuss the geometric structure of spatial curvature and of its
fluctuations in observational cosmology, let us recall the essential steps
required for constructing a cosmological spacetime out of the evolution of 
a Riemannian three--dimensional manifold $\Sigma$, which we assume for
simplicity to be closed and without boundary. Note, however, that such a
condition is not essential for our analysis and in due course it will be
substantially relaxed. The geometry and the matter content of  such a
three--manifold is described by a suitable set of initial data (latin indices run through
$1,2,3$; we adopt the summation convention)
\begin{equation}
(\Sigma \;,\;g_{ab},\;K_{ab},\;\varrho ,\;J_{a})\;,  \label{idata}
\end{equation}
subjected to the energy (Hamiltonian) and momentum (Codazzi) constraints: 
\begin{equation}
{\cal R}+{K}^{2}-K_{\;\,b}^{a}K_{\;\,a}^{b}=16\pi G\varrho +2\Lambda
\;\;;\;\;
\nabla _{b}K_{\;\,a}^{b}-\nabla _{a}K =8\pi GJ_{a}\;,  \label{constraints}
\end{equation}
where $\Lambda$ is the cosmological constant, $K:=g^{ab}K_{ab}$, and where 
${\cal R}$  is the scalar curvature of the Riemannian metric $g_{ab}$;
the covariant spatial derivative with respect to $g_{ab}$ is denoted by $\nabla_a$. 

\begin{figure}[h]
\begin{center}
\includegraphics[bb= 0 0 540 470,scale=.6]{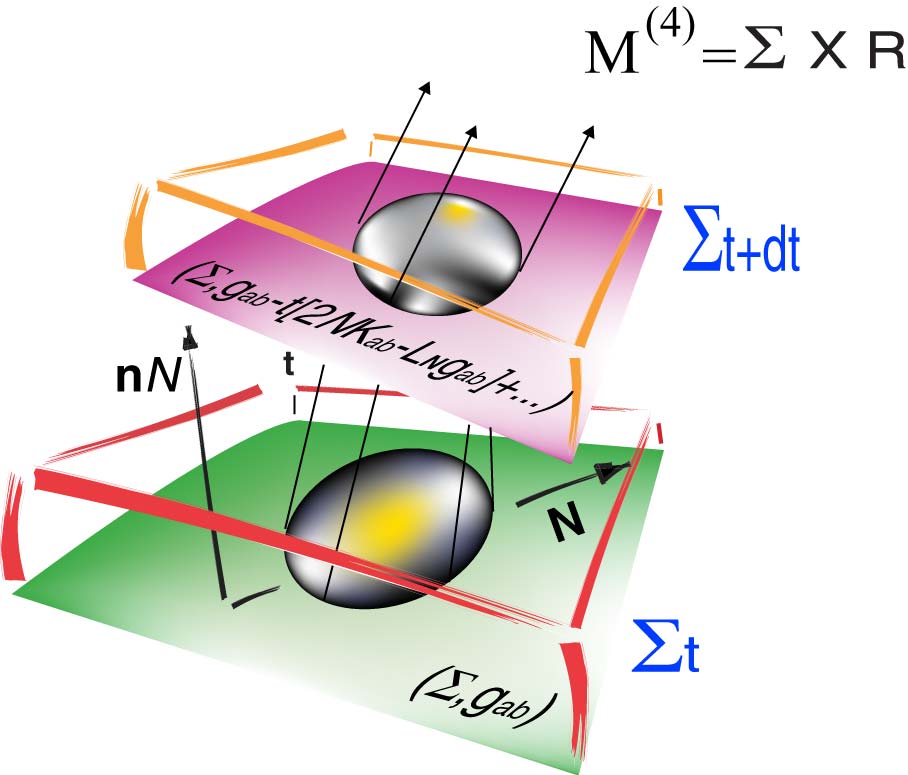}
\caption{The kinematics of the $3+1$--splitting of a cosmological spacetime. The second fundamental form is here represented by the deformation of the shaded domain along the spacetime vector field $\vec{t}=\vec{N}+N\,\vec{n}$, where $\vec{n}$ denotes the normal to $\Sigma_{t}$, and $N$, $\vec{N}$ are the lapse function and  and the shift vector field, respectively.}
\end{center}
\end{figure}

If such a set of admissible data is propagated according to the evolutive part
of Einstein's equations (see Figure 6), then the symmetric tensor field $K_{ab}$ can be
interpreted as the extrinsic curvature (or second fundamental form) of the
embedding $i_{t}: \Sigma \rightarrow M^{(4)}$ of $(\Sigma ,g_{ab})$ in the
spacetime $(M^{(4)}\simeq \Sigma \times \mathbb{R},g^{(4)})$ resulting from
the evolution of $(\Sigma ,g_{ab},K_{ab},\varrho ,J_{a})$, whereas $\varrho$
and $J_{a}$ are, respectively, identified with the mass density and the
momentum density of the material self--gravitating sources on $(\Sigma,g_{ab})$. 
For short we shall call  $(\Sigma ,\;g_{ab},\;K_{ab},\;\varrho,\;J_{a})$ 
the {\em Physical Space} associated with the Riemannian
manifold $(\Sigma ,g_{ab})\hookrightarrow (M^{(4)}\simeq \Sigma \times 
\mathbb{R},g^{(4)})$. In what follows we shall make no use of the evolutive
part of Einstein's equations and accordingly we do not explicitly write it down. 

\vspace{10pt}
\subsection{Averaging scales}
Let us recall the various hierarchical length scales involved that we have described 
in Section~\ref{section:phenomenology}, now associated with the
curvature structure on the physical space $(\Sigma ,g_{ab})$: (i) the
length scale ${L}_\CH$ defined by a spatial region ${\cal D}$ over
which $(\Sigma ,g_{ab})$ can be viewed as describing to a good approximation 
a homogeneous and isotropic state (being not necessarily a
homogeneous--isotropic solution of Einstein's equations); (ii) the
length scales associated with the smaller domains over which the typical
cosmological inhomogeneities regionally dominate, with an alternance of underdense regions 
${\cal E}^{(\alpha )}$ and matter--dominated regions ${\cal M}^{(i)}$,
with ${\cal D}
=\lbrace \cup _{\alpha }{\cal E}^{(\alpha )}\rbrace \cup \lbrace \cup _{i}{\cal M}^{(i)}\rbrace$, 
${\cal E}^{(\alpha )}\cap {\cal M}^{(i)}=\emptyset$, ${\cal E}^{(\alpha )}\cap {\cal E}^{(\beta)}=\emptyset$,
${\cal M}^{(i)}\cap {\cal M}^{(j)}=\emptyset$, for all $\alpha \ne \beta$ and for all $i \ne j$. We
denoted these latter length scales by $L_{{\cal E}}$ and 
$L_{{\cal M}}$ respectively. For the former we
sometimes say simply `voids' or `empty regions', but 
our calculations are kept more general (see Figure 7). 

\begin{figure}[h]
\begin{center}
\includegraphics[bb= 0 0 540 470,scale=.4]{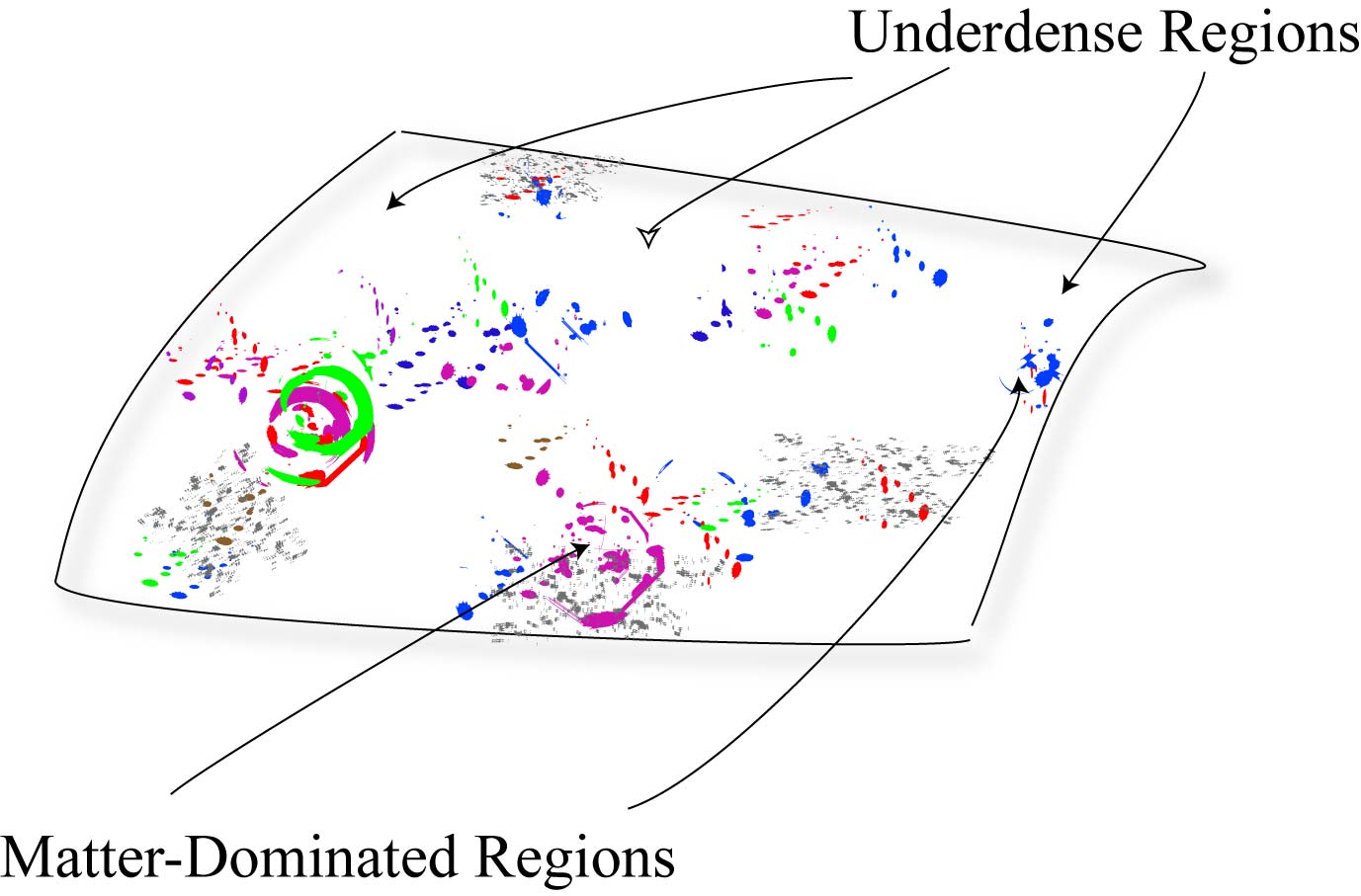}
\caption{The alternance of Underdense and Matter--Dominated regions partitioning the spatial domain $\cal{D}$.}
\end{center}
\end{figure}

\clearpage

In order to discuss the implications generated by such a partitioning of ${\cal D}$, let us rewrite
the Hamiltonian and the momentum constraints (\ref{constraints}) over $(\Sigma ,g_{ab})$ as 
\begin{equation}
{\cal R}=16\pi G\varrho +2\sigma ^{2}-\frac{2}{3}{\theta }^{2}+2\Lambda \;, 
\label{HamR}
\end{equation}
\begin{equation}
\nabla ^{a}\sigma _{ab}=\frac{2}{3}\nabla _{b}\theta -8\pi GJ_{b} \;,
\label{DivCon}
\end{equation}
where ${\theta := -}K_{\;b}^{b}$ is the local {\em rate of expansion}, and 
$\sigma ^{2}:=$ $1/2 \sigma _{\;\,}^{ab}\sigma _{\,ab}$ the square of the local 
{\em rate of shear}, with $\sigma _{\,ab}:= -(K_{ab}-
\frac{1}{3}g_{ab}K)$ being the shear tensor defined by $K_{ab}$.  We wish
to average (\ref{HamR}) over the region ${\cal D}
=\lbrace \cup _{\alpha }{\cal E}^{(\alpha )}\rbrace \cup \lbrace \cup _{i}{\cal M}^{(i)}\rbrace$ and
discuss to what extent such an averaged constraint characterizes the sign of the
curvature. Let us observe that on the scale of near--homogeneity $L_\CH$ the averaged Hamiltonian constraint (\ref{HamR}) is \emph{assumed} to have the structure (\ref{averagehamiltonD})
\begin{equation}
\label{averageham}
{\average{\CR}\;=\; -6H_\CD^2 -{\cal Q}_\CD} + {16\pi G}\langle\varrho\rangle_\CD
+2\Lambda\;,
\end{equation}
where
\begin{equation}
H_{{\cal D}}:= \frac{1}{3}\langle {\theta }\rangle _{{\cal D}}\;,
\end{equation}
is the (average) Hubble parameter on the scale $L_\CH$. This assumption characterizes the kinematical backreaction term 
${\cal Q}_\CD$ as 
\begin{equation}
{\cal Q}_\CD=\frac{2}{3}\langle{\theta }^{2}\rangle_\CD-6\,H_\CD^2-2\langle\sigma ^{2}\rangle_\CD\;.
\end{equation}
Our strategy will be to express both $\langle {\cal R}\rangle_\CD$ and ${\cal Q}_\CD$ in terms of the \emph{typical} local fluctuations of $\varrho$, $\sigma $, and $\theta $ in the voids ${\cal E}^{(\alpha )}$ and in the matter dominated regions ${\cal M}^{(i)}$ (see Figure 8).
As a preliminary step, let us consider the average of a generic scalar--valued
function $f$ over ${\cal D}$,
\begin{equation}
\langle f\rangle _\CD:= | {\cal D}|^{-1}\int_{{\cal D}}f\,d\mu _{g}\;,
\end{equation}
where $| {\cal D}| := $ $\int_{{\cal D}}d\mu _{g}$, and
$d\mu _{g}:= \sqrt{g}dx^{1}dx^{2}dx^{3}$, the Riemannian volume
of ${\cal D}$. If we partition ${\cal D}$ according to ${\cal D}
=\lbrace\cup _{\alpha }{\cal E}^{(\alpha )}\rbrace \cup \lbrace\cup _{i}{\cal M}^{(i)}\rbrace$, where all individual regions
are disjoint in the partitioning, then we can
rewrite $\langle f\rangle _{{\cal D}}$ as 
\begin{equation}
\langle f\rangle _{{\cal D}}=| {\cal D}|
^{-1}\sum_{\alpha }| {\cal E}^{(\alpha )}| \langle
f\rangle _{{\cal E}^{(\alpha )}}+| {\cal D}|
^{-1}\sum_{i}| {\cal M}^{(i)}| \langle f\rangle _{{\cal M}^{(i)}}\;,
\end{equation}
where 
\begin{equation*}
\langle f\rangle _{{\cal E}^{(\alpha )}} := | 
{\cal E}^{(\alpha )}| ^{-1}\int_{{\cal E}^{(\alpha )}}f\,d\mu_{g}\;,\quad
\langle f\rangle _{{\cal M}^{(i)}} := | {\cal M}^{(i)}|^{-1}
\int_{{\cal M}^{(i)}}f\,d\mu _{g}\;. \nonumber
\end{equation*}
\noindent Since both the averages $\langle f\rangle_{{\cal E}^{(\alpha )}}$, 
$\langle f\rangle _{{\cal M}^{(i)}}$ and the 
corresponding regions $| {\cal E}^{(\alpha)}| $ and 
$| {\cal M}^{(i)}| $ may fluctuate in value and size over the set of underdense $\{{\cal E}^{(\alpha )}\}_{\alpha =1,2,...}$  and overdense regions $\{{\cal M}^{(i)}\}_{i=1,2,....}$, it is  useful to introduce the weighted averages of $\langle f\rangle_{{\cal E}^{(\alpha )}}$ and of $\langle f\rangle _{{\cal M}^{(i)}}$, \emph{viz.}
\begin{equation}  
\langle
f\rangle _{{\cal E}}:=\frac{\sum_{\alpha }\,| {\cal E}^{(\alpha)}|\,\langle f\rangle _{{\cal E}^{(\alpha )}}}{\sum_{\beta }\,| {\cal E}^{(\beta )}|}\,=\,| {\cal D}_{{\cal E}}|^{-1}\,\int_{\cup_{\alpha } {\cal E}^{(\alpha)}}f\,d\mu _{g} \;,
\end{equation}
and 
\begin{equation}  
\langle
f\rangle _{{\cal M}}:=\frac{\sum_{i }\,| {\cal M}^{(i)}|\,\langle f\rangle _{{\cal M}^{(i )}}}{\sum_{k }\,| {\cal M}^{(k )}|}\,=\,| {\cal D}_{{\cal M}}|^{-1}\,\int_{\cup_{i } {\cal M}^{(i)}}f\,d\mu _{g}\;,
\end{equation}
where $| {\cal D}_{{\cal E}}| := \sum_{\alpha}| {\cal E}^{(\alpha )}
|$, and $| {\cal D}_{{\cal M}}| :=\sum_{i}| {\cal M}^{(i)}|$. Since ${\cal D}
=\lbrace\cup _{\alpha }{\cal E}^{(\alpha )}\rbrace \cup \lbrace\cup _{i}{\cal M}^{(i)}\rbrace$,\,\, $| {\cal D}_{{\cal E}}| +$\ $| {\cal D}_{{\cal M}}| =
| {\cal D}| $, we have
\begin{equation}
\langle f\rangle _{{\cal D}}=\frac{| {\cal D}_{{\cal E}}| }
{| {\cal D}| }\langle f\rangle_{{\cal E}}+\frac{| {\cal D}_{{\cal M}}
| }{| {\cal D}| }\langle f\rangle _{{\cal M}}\;.
\label{splitAv}
\end{equation}
If we now introduce the adimensional parameter 
\begin{equation}
\lambda _{{\cal M}}:= \frac{| {\cal D}_{{\cal M}}| }{| {\cal D}|}\;,
\end{equation}
we can write (\ref{splitAv}) equivalently as 
\begin{equation}
\langle f\rangle _{{\cal D}}=(1-\lambda _{{\cal M}})\,
\langle f\rangle _{{\cal E}}+\lambda _{{\cal M}}\,
\langle f\rangle _{{\cal M}}\;.
\end{equation}

\begin{figure}[h]
\begin{center}
\includegraphics[bb= 0 0 540 470,scale=.4]{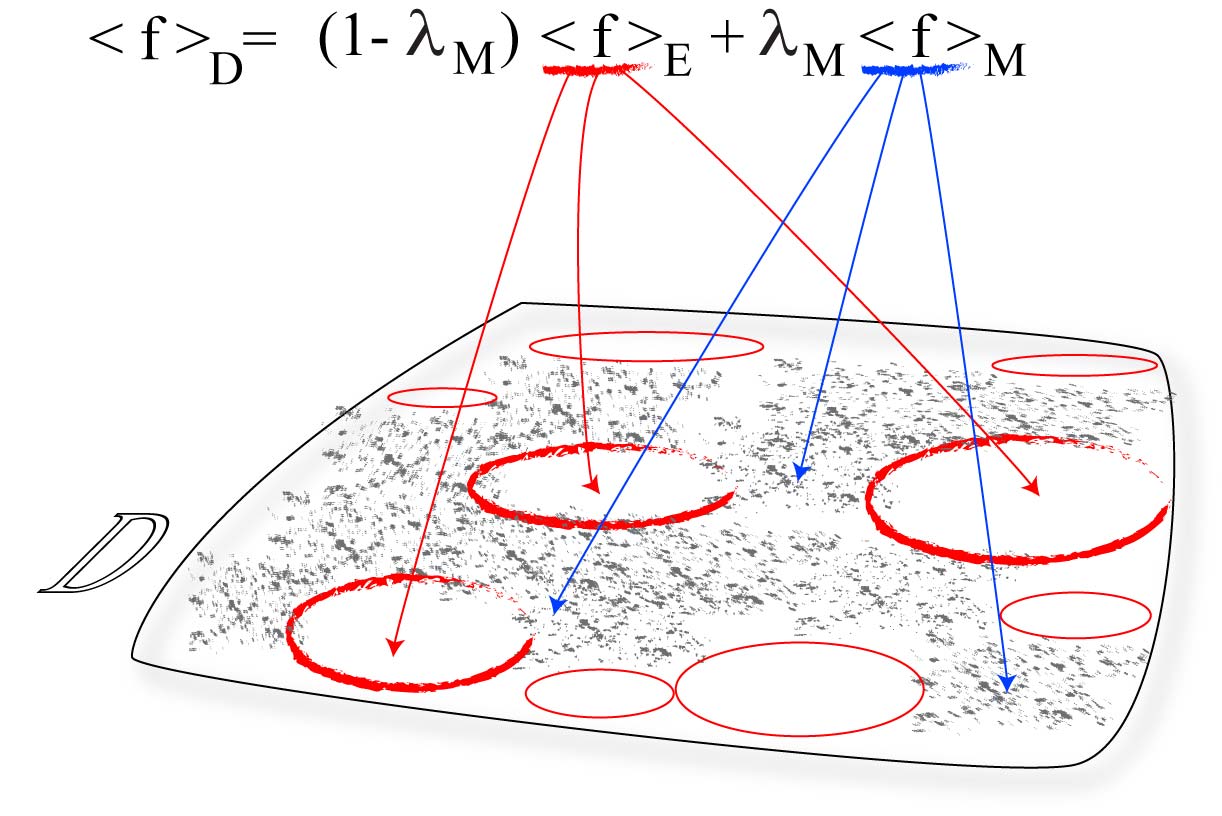}
\caption{Splitting the average of the function $f$ over the spatial region $\cal{D}$ into the weighted contributions coming from the underdense and from the matter--dominated regions.}
\end{center}
\end{figure}

Applying in turn this formula to the volume--average of the scalar curvature ${\cal R}$,
then we simply get 
\begin{equation}
\langle {\cal R}\rangle _{{\cal D}}=(1-\lambda _{{\cal M}})\,
\langle {\cal R}\rangle _{{\cal E}}+\lambda _{{\cal M}}\,
\langle {\cal R}\rangle _{{\cal M}}\;,
\end{equation}
which, according to (\ref{HamR}) implies
\begin{eqnarray*}
\langle {\cal R}\rangle _{{\cal D}}-2\Lambda \\
\fl
= (1-\lambda _{{\cal M}})
\left[ 16\pi G\langle \varrho 
\rangle_{{\cal E}}+2\langle \sigma ^{2}\rangle _{{\cal E}}-\frac{2}{3}
\langle {\theta }^{2}\rangle _{{\cal E}}\right]
+\lambda _{{\cal M}}\left[ 16\pi G\langle \varrho 
\rangle_{{\cal M}}+2\langle \sigma ^{2}\rangle _{{\cal M}}-
\frac{2}{3}\langle {\theta }^{2}\rangle _{{\cal M}}\right]
\;.
\end{eqnarray*}

\subsection{The Kinematical Backreaction}
\label{Hubble}
At this stage, we can look at the regional Hubble parameters $H_{{\cal E}}$ and 
$H_{{\cal M}}$ assigned to the empty and
matter--dominated regions and their associated mean--square fluctuations 
$\delta ^{2}H_{{\cal E}}$ and $\delta ^{2}H_{{\cal M}}$ according to 
\begin{eqnarray}
H_{{\cal E}}:= \frac{1}{3}\langle {\theta }\rangle _{{\cal E}}\;\;;\;\;
H_{{\cal M}}:= \frac{1}{3}\langle {\theta }\rangle _{{\cal M}}\;\;; \nonumber\\
\delta ^{2}H_{{\cal E}}:= \frac{1}{9}\left( \langle {\theta }^{2}
\rangle_{{\cal E}}-\langle {\theta }\rangle _{{\cal E}}^{2}\right) 
\;\;;\;\;\delta ^{2}H_{{\cal M}}:= \frac{1}{9}
\left( \langle {\theta }^{2}\rangle _{{\cal M}}-\langle 
{\theta }\rangle _{{\cal M}}^{2}\right)\;. 
\end{eqnarray}
One easily computes 
\begin{equation}
\fl
\langle {\cal R}\rangle _{{\cal D}}-2\Lambda=(1-\lambda _{\cal{M}})
\left[ 16\pi G\langle
\varrho \rangle _{{\cal E}}-6H_{{\cal E}}^{2}-{\cal Q}_\CE \right]
+\lambda _{{\cal M}}\left[ 16\pi G\langle
\varrho \rangle _{{\cal M}}-6H_{{\cal M}}^{2}-{\cal Q}_\CM\right],
\label{R}
\end{equation}
where ${\cal Q}_\CE$ and ${\cal Q}_\CM$ denote the {\em kinematical backreaction terms}
on the respective scales:
\begin{equation}
{\cal Q}_\CE := 6\,\delta ^{2}H_\CE - 2\langle \sigma^{2}\rangle _\CE
\end{equation}
 and 
\begin{equation}
{\cal Q}_\CM := 6\,\delta ^{2}H_\CM - 2\langle \sigma^{2}\rangle _\CM.
\end{equation}
If we insert (\ref{R}) into the expression (\ref{averageham}) characterizing ${\cal Q}_\CD$
we get
\begin{equation}
{\cal Q}_\CD = (1-\lambda_\CM ){\cal Q}_\CE + \lambda_\CM {\cal Q}_\CM +6(1-\lambda_\CM )\,H^{2}_{\CE}
+6\lambda_\CM\,H^{2}_{\CM}-6H^{2}_{\CD}\;.
\end{equation}
Since
\begin{equation}
H_{\CD}=(1-\lambda_\CM )\,H_{\CE}
+\lambda_\CM\,H_{\CM}\;,
\label{Hfactor}
\end{equation}
a direct computation provides
\begin{equation}
{\cal Q}_\CD = (1-\lambda_\CM ){\cal Q}_\CE + \lambda_\CM {\cal Q}_\CM 
+6{\lambda_\CM}(1-\lambda_\CM)\,\left(H_\CE-H_\CM\right)^{2} \;,
\end{equation}
or, more explicitly,
\begin{equation}
\fl
{\cal Q}_\CD = 6(1-\lambda_\CM )\,\delta ^{2}H_\CE + 6\lambda_\CM\,\delta ^{2}H_\CM 
+6{\lambda_\CM}(1-\lambda_\CM)\,\left(H_\CE-H_\CM\right)^{2}-2\,\langle\sigma ^{2}\rangle_\CD \;.
\label{mexpl}
\end{equation}
The above formulae for the averaged curvature and kinematical backreaction are {\em general} for our choice of a partitioning into {\em overdense} and {\em underdense} domains. 

\vspace{10pt}

It is important to observe that in the factorization (\ref{Hfactor}) both $H_{\CE}$ and $H_{\CM}$ are effectively functions of $\lambda_{\CM}$, 
(this is simply a fact coming from the definition of the average factorization we have used), and for discussing the meaning of the expressions we obtained for the kinematical backreaction ${\cal Q}_\CD$ and scalar curvature $\langle {\cal R}\rangle _{{\cal D}}$ it is often useful to assume a reasonable scaling for 
$H_{\CE}(\lambda_{\CM})$ and $H_{\CM}(\lambda_{\CM})$. Our basic understanding is that, on small scales, the local dynamics of gravitationally bound matter will obliterate $H_{\CM}$, whereas, if matter happens to be distributed over larger and larger domains, then it will more and more participate in the global averaged dynamics. By continuity, there should be a scale $\lambda^{cr}_\CM$ marking a significant transition between these two regimes. Clearly, one can elaborate on the most appropriate model for such a transition, but the one described below, basically a Gaussian modeling, is quite general and has the merit of avoiding sudden jumps in the behavior of $H_{\CM}$. Also, it can be a natural starting point for a more elaborate analysis. Thus, we wish to make an
idealization by assuming the ``stable--clustering hypethesis'' to hold on the matter--dominated regions, 
$H_\CM \cong 0$, it will only hold up to a critical scale $\lambda^{cr}_\CM\in(0,1)$, 
whereas for $\lambda^{cr}_\CM\leq\lambda_\CM\leq 1$ the quantity $H_{\CM}(\lambda_{\CM})$ smoothly increases up to $H_{\CD}$. 
To achieve this we model the scale--dependence of $H_{\CM}(\lambda_{\CM})$  according to
\begin{equation} 
\fl\qquad
H_{\CM}(\lambda_{\CM}):= H_{\CD}\,\exp\left[-\frac{(\lambda_\CM-1)^{2}}{(\lambda^{cr}_\CM-1)^{2}-(\lambda_\CM-1)^{2}} \right]
\quad {\rm for}\quad \lambda^{cr}_\CM\leq\lambda_\CM\leq 1\;\;,
\label{scalingmodel1}
\end{equation}
and
\begin{equation}
\fl\qquad
H_{\CM}(\lambda_{\CM}):= 0 \quad {\rm for} \quad 0\leq \lambda_{\CM} \leq \lambda^{cr}_\CM \;\;.
\end{equation}
It is easily verified that $H_{\CM}(\lambda_{\CM})$ is a smooth ($C^{\infty }$) function of $\lambda_{\CM}\in [0,1]$  with 
support in $\lambda^{cr}_\CM\leq\lambda_\CM\leq 1$, and which vanishes together with all its derivatives at $\lambda_\CM=\lambda^{cr}_\CM$. 
(The graph of $H_{\CM}(\lambda_{\CM})$ (see Figure 9) is the left--half of a bell--shaped function, smoothly rising from zero at $\lambda_\CM=\lambda^{cr}_\CM$, and reaching its maximum at $\lambda_\CM=1$).

\begin{figure}[h]
\begin{center}
\includegraphics[bb= 0 0 540 470,scale=.4]{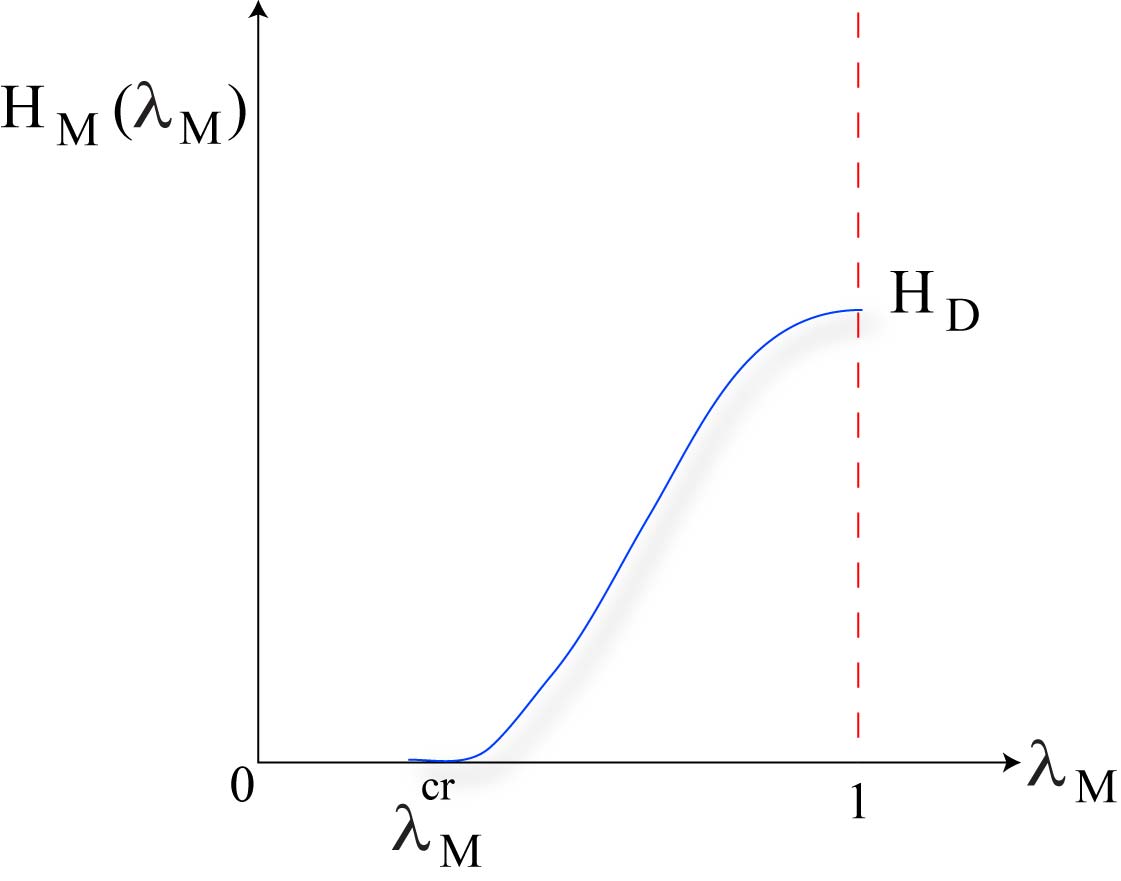}
\caption{The graph of $H_{\CM}(\lambda_{\CM})$.}
\end{center}
\end{figure}

From the factorization $H_{\CD}=(1-\lambda_\CM )\,H_{\CE}+\lambda_\CM\,H_{\CM}$ we then get
\begin{equation}
\fl\qquad
H_{\CE}(\lambda_{\CM}):=\frac{H_{\CD}}{1-\lambda_{\CM}}\,\left({1-\lambda_{\CM}\,e^{-\frac{(\lambda_\CM-1)^{2}}{(\lambda^{cr}_\CM -1)^{2}-(\lambda_\CM-1)^{2}} }}\right)\quad {\rm for} \quad \lambda^{cr}_\CM \leq\lambda_\CM\leq 1 \;,
\end{equation}
and 
\begin{equation}
\fl\qquad
H_{\CE}(\lambda_{\CM}):= \frac{H_{\CD}}{1-\lambda_{\CM}} \quad {\rm for} \quad 0\leq \lambda_{\CM} \leq \lambda^{cr}_\CM \;\;.
\end{equation}
\smallskip

In light of the above remarks,  we can, for $0\leq \lambda_{\CM} \leq \lambda^{cr}_\CM$ , consider the idealization  $H_\CD \cong (1-\lambda_\CM )H_\CE$ and $H_\CM \cong 0$, which together with 
 $\langle\varrho\rangle_\CE \cong 0$ and 
$\langle\varrho\rangle_\CD \cong \lambda_\CM \langle\varrho\rangle_\CM$, allows to write  the total kinematical backreaction in the simpler form:
\begin{equation}
\label{idealizedQ}
{\cal Q}_\CD = (1-\lambda_\CM ){\cal Q}_\CE + \lambda_\CM {\cal Q}_\CM 
+\frac{\lambda_\CM}{1-\lambda_\CM}\,6 H_\CD^2 \;.
\end{equation}

\vspace{10pt}

Thus, if we consider the fluctuations in the regional expansion rates $\delta ^{2}H_\CE$ and $\delta ^{2}H_\CM$, the effective regional expansion rates themselves,
as well as  the regional matter contents $\Omega^\CE_m$ and $\Omega^\CM_m$, and the volume fraction of matter 
$\lambda_\CM$, as in principle observationally determined quantities\footnote{Of course, the interpretation of these observations would still be
model--dependent, see the discussion in Sect.~\ref{section:conclusions}.},
our main task will be to estimate the shear terms in $\Omega^\CD_{\cal Q}$ from its regional contributions in 
order to find an estimate for the curvature parameter. 
We shall now turn to this problem and investigate the shear terms that contain matter shear and geometrical shear and are therefore, like the
intrinsic curvature, not directly accessible through observations; thus, we aim at 
deriving estimates that trace the shear terms back to the above more accessible parameters.

\clearpage

\section{Estimating kinematical backreaction}
\label{section:Q}

To provide control on ${\cal Q}_\CD$ we need to better understand the
relative importance of the shear term $\langle \sigma
^{2}\rangle _{{\cal D}}$. Notoriously, this quantity is difficult to estimate without referring to particular models for the matter and 
gravitational radiation distributions, and in cases where we do not describe the variables in a comoving frame. In order to avoid a specific modelling and a specific coordinate choice,  our strategy here is to express $\langle \sigma^{2}\rangle _{{\cal D}}$ in terms of the correlation properties of the shear--generating currents. These correlation properties characterize the  length scales over which the shear can be physically significant and allow for a natural parametrization of $\langle \sigma^{2}\rangle _{{\cal D}}$. Also, the underlying rationale is to separate, in each underdense and overdense region $\cal E$ and $\cal M$, the contribution to the shear coming from the local dynamics of matter and expansion, and to
isolate this from the contribution coming from cosmological gravitational radiation. 
Technically, this procedure is based on a decomposition of the shear into  its longitudinal (matter--expansion) and transverse (radiation) parts with respect to the  $L^{2}(\Sigma  )$ 
inner product defined  by 
\begin{equation}
(W,V)_{L^{2}(\Sigma  )}:= \int_{\Sigma  }g^{il}g^{km}W_{ik}V_{lm}\,d\mu _{g}\;,
\end{equation}
where $W_{ik}$, $V_{lm}$\ are (square--summable) symmetric bilinear forms compactly supported in $\Sigma $.
With this remark along the way, we start by observing that for a given $\sigma _{ab}$ we can write 
\begin{equation}
\sigma _{ab}=\sigma _{\perp ab}+\sigma _{\parallel ab}\;,  \label{Ddecomp}
\end{equation}
with 
\begin{equation}
\left( \sigma _{\perp },\sigma _{\parallel }\right) _{L^{2}(\Sigma  )}=0\;,
\label{Ldue}
\end{equation}
where $\sigma _{\perp ab}$ and $\sigma _{\parallel ab}$ respectively are
the divergence--free part and the longitudinal part of $\sigma _{ab}$ in  $\Sigma $. 

\begin{figure}[h]
\begin{center}
\includegraphics[bb= 0 0 540 470,scale=.4]{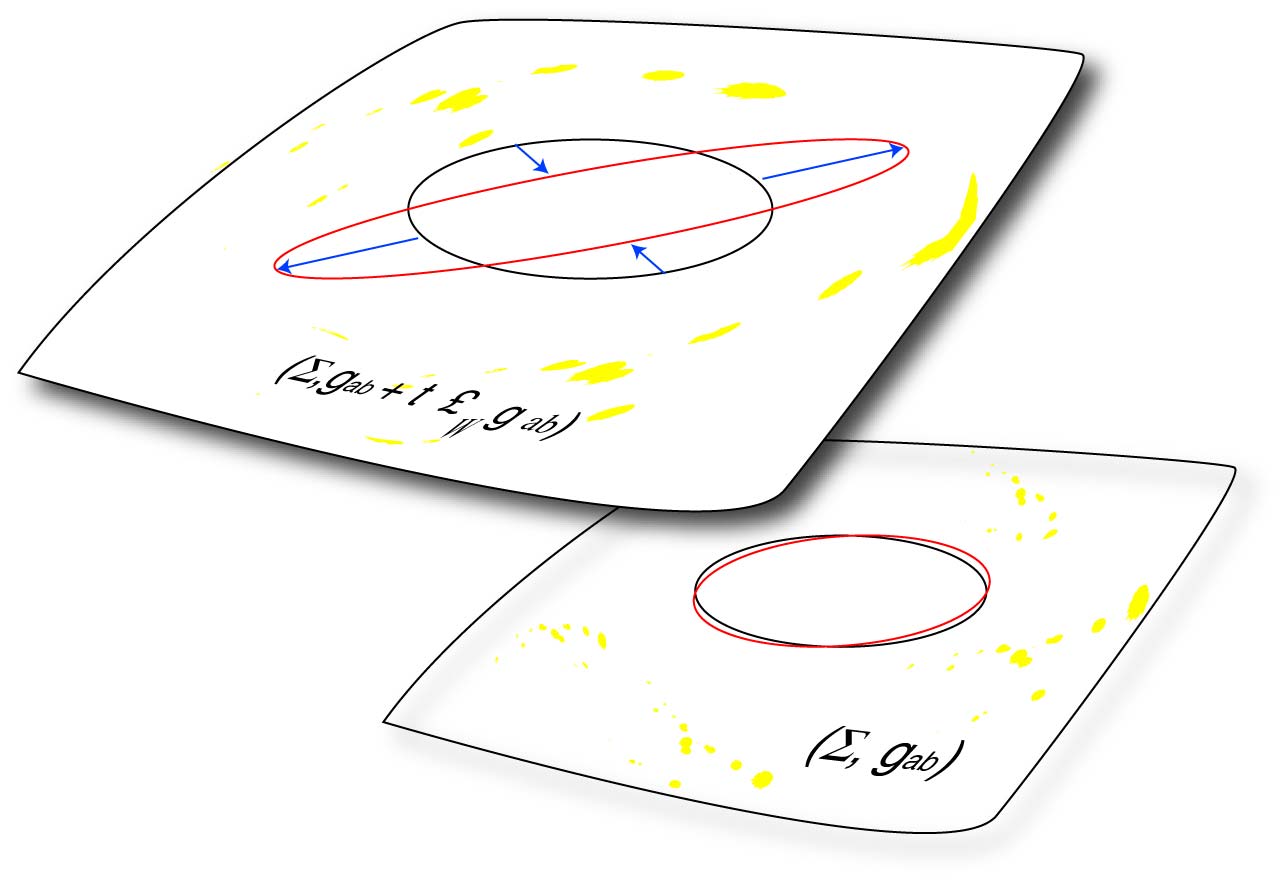}
\caption{The longitudinal shear $\sigma _{\parallel ab}$ associated with a 1--parameter family of  volume--preserving diffeomorphisms in a fixed background geometry. The shearing diffeomorphisms are generated by a vector field $\vec{w}$, solution of the elliptic equation (\ref{vectW}), with sources given by $\frac{2}{3}\nabla _{b}\theta -8\pi GJ_{b}$.}
\end{center}
\end{figure}

Recall that $\sigma _{\perp ab}$ generates non--trivial (proper time--)deformations of the conformal geometry
associated with $g_{ab}$, and represents the dynamical part of the shear, associated with
the presence of gravitational radiation in $\Sigma  $. Conversely, the term $\sigma _{\parallel ab}$
is a gauge term associated with the infinitesimal ${\rm Diff}(\Sigma  )$--reparametrization of the
conformal geometry generated by the motion of matter and inhomogeneities in the gradient of the expansion rate.
The $L^{2}(\Sigma  )-$orthogonality (\ref{Ldue}) of these two terms  
implies that 
\begin{equation}
\int_{\Sigma  }\sigma ^{2}d\mu _{g}=\int_{\Sigma }\sigma _{\perp }{}^{2}d\mu
_{g}+\int_{\Sigma  }\sigma _{\parallel }{}^{2}d\mu _{g}\;,
\end{equation}
where $\sigma _{\perp }{}^{2}:= 1/2 g^{ab}g^{cd}\sigma _{\perp ac}\sigma _{\perp bd}$ 
and  
$\sigma_{\parallel }{}^{2}:= 1/2 g^{ab}g^{cd}\sigma _{\parallel ac}\sigma_{\parallel bd}$.
Explicitly, the terms $\sigma _{\perp ab}$ and $\sigma _{\parallel ab}$ (see Figures 10 and 11) are characterized by
\begin{eqnarray}
\nabla ^{a}\sigma _{\perp ab}=0 \\
\sigma _{\parallel ab}=\nabla _{a}w_{b}+\nabla _{b}w_{a}-\frac{2}{3}
g_{ab}\nabla ^{c}w_{c}:= {\pounds }_{\vec w}\,g_{ab}\;,
\end{eqnarray}
where ${\pounds }_{\vec w}\,g_{ab}$ denotes the conformal Lie
derivative of the metric in the direction of the vector field $w^{a}$.
This latter is the solution (unique up to conformal Killing vectors--see below) of the
elliptic partial differential equation $\nabla ^{a}
\left( {\pounds }_{\vec w}g_{ab}\right) =\nabla ^{a}\sigma _{ab}$, \emph{i.e.} 
\begin{equation}
\Delta w_{b}+\frac{1}{3}\nabla _{b}(\nabla _{a}w^{a})+{\cal R}_{ab}w^{a}=
\nabla ^{a}\sigma _{ab}\;.
\label{vectWdef}
\end{equation}

\begin{figure}[h]
\begin{center}
\includegraphics[bb= 0 0 540 470,scale=.4]{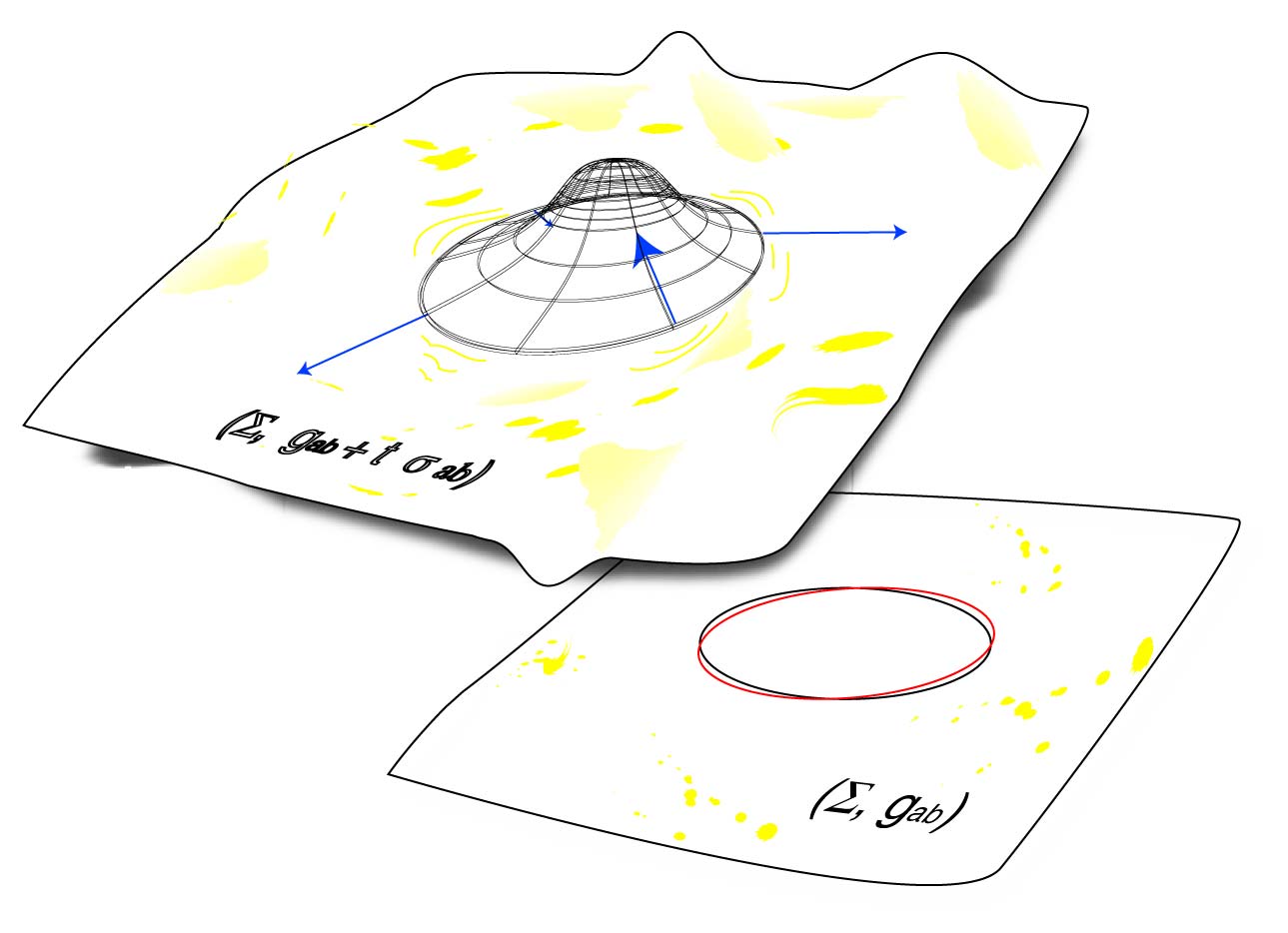}
\caption{The transverse shear $\sigma _{\perp ab}$ associated with a volume--preserving (infinitesimal) deformation of the  background metric geometry of $\Sigma$.}
\end{center}
\end{figure}
If we take into account the momentum constraints (\ref{DivCon}), 
$\nabla ^{a}\sigma _{ab}=\frac{2}{3}\nabla _{b}\theta -8\pi GJ_{b}$,
the elliptic equation determining the vector field $w_{b}$ becomes 
\begin{equation}
\Delta ^{a}_{\;b}\;w_{a}=\frac{2}{3}\nabla _{b}\theta -8\pi GJ_{b}\;,
  \label{vectW}
\end{equation}
where 
\begin{equation}
\Delta ^{a}_{\;b}:=
\Delta \delta^{a}_{\;b}+\frac{1}{3}\nabla ^{a}\nabla _{b}+{\cal R}^{a}_{\;b}
\end{equation}
is the formally self--adjoint elliptic vector Laplacian on the space of vector fields on $(\Sigma ,g_{ab})$.

\subsection{The region of near--homogeneity $\mathcal{D}$}

 In order to characterize geometrically the region of near--homogeneity $\mathcal{D}$ let us recall some basic properties of the vector Laplacian $\Delta ^{a}_{\;b}$ introduced above.  As $x$ varies over $(\Sigma, g)$, let $\{\vec{\psi }_{(i)}(x)\}_{i=1,2,3}$ denote  a local frame in $T_{x}\Sigma $. For later convenience, we normalize the vector fields $x\mapsto \vec{\psi }_{(i)}(x)$ according to $\int_{\Sigma }{\psi }^{a}_{(i)}\,{\psi }_{a}^{(k)}\,d\mu _{g}\,=\,|\Sigma |\,\delta _{i}^{k}$, and, when necessary, we  assume that $\{\vec{\psi }_{(i)}(x)\}_{i=1,2,3}$ is an orthonormal frame in $T_{x}\Sigma $.  For each given  $x\mapsto \vec{\psi }_{(i)}(x)$, let $\vec{e }_{(i)}(x)$ be the corresponding vector field defined by the action of the operator $\Delta ^{a}_{\;b}$, \emph{i.e.},
\begin{equation}
C_{(i)}^{-1}\;e^{(i)}_{b}:=\Delta ^{a}_{\;b}\;\psi ^{(i)}_{a}\;,
\label{defeC}
\end{equation}
where $C_{(i)}$ is a normalization constant  (with the dimension of a length squared) that will be chosen momentarily. We denote by
$\sigma^{(i)} _{ab}:={\pounds }_{\vec{\psi }_{(i)} }\,g_{ab}$ the shear associated with $\vec{\psi }_{(i)}(x)$.
Observing  that $ \sigma^{(i)} _{ ab}\,\nabla ^{a}
\psi _{(i)}^{b}$ $=$ $\frac{1}{2}\,\sigma^{(i)} _{ ab}\,\sigma _{(i)}^{\,ab}$, (no summation over $(i)$), and integrating by parts we get
\begin{equation}
\int_{\Sigma }\psi ^{b}_{(i)}\nabla ^{a}(\sigma^{(i)} _{ab})\,d\mu _{g}=-\frac{1}{2}\,
\int_{\Sigma  }\sigma _{(i)}{}^{2}\,\,d\mu _{g}\;,
\label{sfrutto}
\end{equation}
which, if we exploit the relation $\nabla ^{a}(\sigma^{(i)} _{ab})=\Delta ^{a}_{\;b}\;\psi ^{(i)}_{a}$, (see \ref{vectWdef}), yields
\begin{equation}
\int_{\Sigma  }\sigma _{(i)}{}^{2}\,\,d\mu _{g}=\,2\,\int_{\Sigma  } \psi ^{b}_{(i)}\,\left(- \Delta ^{a}_{\;b}\right)\;\psi ^{(i)}_{a}\,\,d\mu _{g}  \;.
\label{positivity}
\end{equation}
This latter  shows that $\left(- \Delta ^{a}_{\;b}\right)$ is a positive operator whose kernel
is generated by the conformal Killing vectors $\{\vec{\xi }_{(\alpha )}\}$ of $(\Sigma ,g_{ab})$. By proceeding similarly, from the relation 
$ \sigma^{(k)} _{ ab}\,\nabla ^{a}
\psi _{(i)}^{b}$ $=$ $\frac{1}{2}\,\sigma^{(k)} _{ ab}\,\sigma _{(i)}^{\,ab}$, with $i\not=k$, we get
\begin{equation}
\int_{\Sigma }\psi ^{b}_{(i)}\nabla ^{a}(\sigma^{(k)} _{ab})\,d\mu _{g}=-\frac{1}{2}\,
\int_{\Sigma  }\sigma^{(k)} _{ ab}\,\sigma _{(i)}^{\,ab}\,\,d\mu _{g}\;,
\label{sfruttoik}
\end{equation}
and
\begin{equation}
\int_{\Sigma  }\sigma^{(k)} _{ ab}\,\sigma _{(i)}^{\,ab}\,\,d\mu _{g}=\,2\,\int_{\Sigma  } \psi ^{b}_{(i)}\,\left(- \Delta ^{a}_{\;b}\right)\;\psi ^{(k)}_{a}\,\,d\mu _{g}  \;.
\label{iktivity}
\end{equation}
In the absence of local conformal symmetries, which we assume to be generically the case in our setting, the left member of (\ref{positivity}) is different from zero, and (\ref{defeC}) can be interpreted as the statement that, if there are no conformal Killing vectors, then there always exists a current $C_{(i)}^{-1}\,\vec{e}_{(i)}$ generating the given shear vector potential $\vec{\psi}_{(i)}$. Moreover  from $\nabla ^{a}(\sigma^{(i)} _{ab})=C_{(i)}^{-1}\,{e}_{a}^{(i)}$  and (\ref{sfrutto}) we get (dividing by $Vol(\Sigma ,g)$),
\begin{equation}
\left\langle\psi ^{b}_{(i)}\,e^{(i)} _{b}\right\rangle_{\Sigma }=-\frac{1}{2}\,C_{(i)}\,
\left\langle\sigma _{(i)}{}^{2}\right\rangle_{\Sigma }\;.
\label{sfruttoC}
\end{equation}
Note that $C_{(i)}\,\langle \sigma _{(i)}{}^{2} \rangle_{\Sigma }$ is a natural measure of the characteristic length scale over which the conformal part of the metric $g^{-\frac{1}{3}}\,g_{ab}$, (a tensor density of weight $-\frac{2}{3}$), changes along the vector field $\vec{\psi } _{(i)}(x)$.
 One should also remark that the ratio 
\begin{equation}
\frac{\int_{\Sigma  }\sigma^{(k)} _{ ab}\,\sigma _{(i)}^{\,ab}\,\,d\mu _{g}}{\left[\langle \sigma _{(i)}{}^{2} \rangle_{\Sigma }\right]^{-\frac{1}{2}}\,\left[\langle \sigma _{(k)}{}^{2} \rangle_{\Sigma }\right]^{-\frac{1}{2}}}\;,
\label{anisoratio}
\end{equation}
provides, for $i\not=\,k$, the typical correlation among the directional shears $\{\sigma^{(k)} _{ ab}\}_{k=1,2,3}$ as the reference direction varies in  $\{\vec{\psi } _{(i)}\}_{i=1,2,3}$. Even if  the manifold $(\Sigma ,g)$ does not admit conformal symmetries, (since we assumed $\ker \Delta ^{a}_{b}=\emptyset$), we should be able to geometrically select  in $(\Sigma ,g)$ a sufficiently large domain $\CD$ of near--homogeneity and near--isotropy. To this end, let us  assume that in $(\Sigma ,g)$ there is a region (containing a sufficiently large metric ball) in which the local orthonormal frame  $x\mapsto \{\vec{\psi } _{(i)}(x)\}$ can be so chosen as to make (\ref{anisoratio}) vanish, and the ratios 
\begin{equation}
\frac{\int_{\Sigma  }\sigma _{(i)}{}^{2}\,\,d\mu _{g}}{\int_{\Sigma  } \psi ^{a}_{(i)}\,\psi ^{(i)}_{a}\,\,d\mu _{g} }
=\left\langle \sigma _{(i)}{}^{2} \right\rangle_{\Sigma }\;,
\label{smallshear}
\end{equation}
(where, in the latter expression, we have exploited the normalization of $\{\vec{\psi } _{(i)}\}$), are as small as possible
with
\begin{equation}
\int_{\Sigma  }\sigma _{(1)}{}^{2}\,\,d\mu _{g}=\int_{\Sigma  }\sigma _{(2)}{}^{2}\,\,d\mu _{g}=\int_{\Sigma  }\sigma _{(3)}{}^{2}\,\,d\mu _{g}\;.
\end{equation}
In order to extract a natural geometric characterization of the region $\CD $ from these requirements let us recall that 
the first positive eigenvalue of the operator $-\Delta ^{a}_{\;b}$ is provided by
\begin{equation}
\lambda _{1}:=\inf \left\{\frac{\int_{\Sigma  } \phi ^{b}\,\left(- \Delta ^{a}_{\;b}\right)\;\phi_{a}\,\,d\mu _{g}  }{\int_{\Sigma  } \phi ^{a}\,\phi_{a}\,\,d\mu _{g}  } \right\}\;,
\end{equation}
where the $\inf$ is over all square--summable vector fields $\vec\phi $ over $(\Sigma ,g)$ which are $L^{2}(\Sigma )$--orthogonal to $\ker\,\Delta ^{a}_{\;b}$. Since from (\ref{smallshear}) we have  
\begin{equation}
\left\langle \sigma _{(i)}{}^{2} \right\rangle_{\Sigma } \geq \; \lambda _{1}\;,\;\;\;\;i=1,2,3\;,
\end{equation}
this naturally suggests to select the local orthonormal frame $\{\vec{\psi } _{(i)}\}$ by choosing the vector fields $\vec{\psi } _{(i)}$ as three independent eigenvectors, without nodal surfaces, of the operator  $-\Delta ^{a}_{\;b}$ associated with the first positive eigenvalue $\lambda _{1}$, \emph{i.e.},
\begin{equation}
-\Delta ^{a}_{\;b}\,{\psi }^{b} _{(i)}=\,\lambda _{1}\,{\psi }^{a} _{(i)}\;,\;\;\;\;\int_{\Sigma }{\psi }^{a}_{(i)}\,{\psi }_{a}^{(k)}\,d\mu _{g}\,=\,|\Sigma |\,\delta _{i}^{k}\;,\;\;\; i,k=1,2,3\;.
\end{equation}
(we are here assuming that the multiplicity of $\lambda_{1}$ is $\geq 3$; for comparison, recall that the multiplicity of the first eigenvalue of the vector Laplacian on the standard $3$--sphere is $6$). According to  (\ref{defeC}),  with such a choice we have $\vec{e}_{(i)}=-\lambda_{1}\,\vec{\psi}_{(i)}$ where the normalization constant takes the natural value $C_{(i)}^{-1}:=-\lambda_{1}$. These remarks allow 
to geometrically characterize the region of near-homogeneity as a region $\left({\CD }, \{\vec{\psi}^{(i)}(x)\}\right)\subseteq (\Sigma ,g)$, of size 
\begin{equation}
 |\CD |:=\, \lambda_{1}^{-\frac{3}{2}}\;,
\label{condscale}
\end{equation}
endowed with the \emph{minimal shear frame} $x\mapsto \{\vec{\psi}^{(i)}(x)\}_{i=1,2,3}$ generated by the chosen three independent nodal--free eigenvectors  associated with  $\lambda _{1}$ (see Figure 12).

\begin{figure}[h]
\begin{center}
\includegraphics[bb= 0 0 540 470,scale=.5]{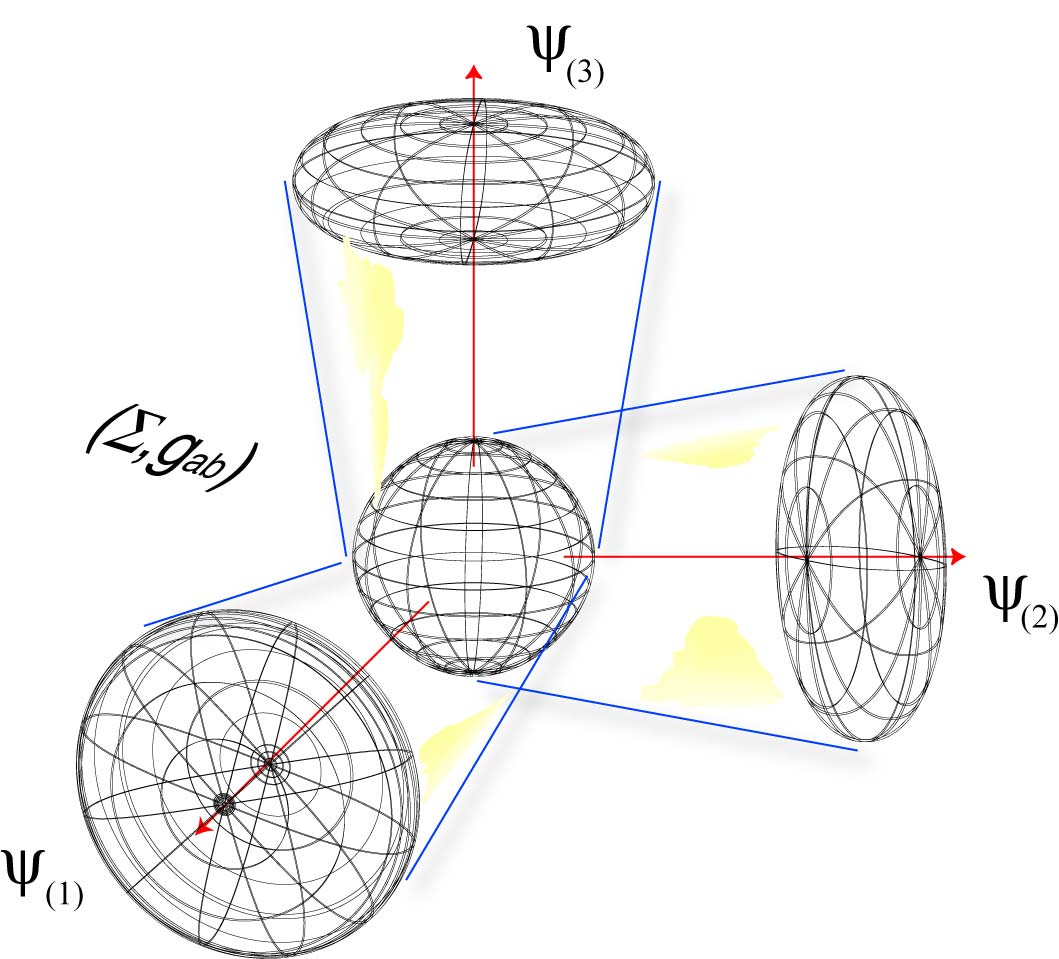}
\caption{A pictorial representation of the (longitudinal) shear generated on a small sphere by three (independent) eigenvectors $\{\vec{\psi}_{(i)}\}_{i=1,2,3}$, associated with the first (degenerate) eigenvalue $\lambda_{1}$ of the vector Laplacian $\Delta ^{a}_{h}$. Such eigenvectors define a moving frame in which the shear is, in a $L^{2}$--average sense, minimized.  The region of near--homogeneity $\mathcal{D}$ can be defined as the largest region in $(\Sigma,g)$ where we can introduce such minimal shear eigenvectors.}
\end{center}
\end{figure}

\subsection{The shear correlation lengths}
\label{shcorrlen}
We can infer a useful consequence of this geometrical characterization of $\CD $ if we exploit the Green function 
$E^{a}_{k'}(x,x')$, of pole 
$x\in \Sigma $, for the vector Laplacian
$\Delta ^{a}_{\;b}$ (see Figures 13 and 14). Let $U\subset \cal D$ be a geodesically convex neighborhood 
containing the point $x$, (we can naturally restrict our attention to the region of near--homogeneity $\CD \subset \Sigma$). For any other point $x'\in U$ let $l(x,x')$ denote the unique 
geodesic segment connecting $x$ to $x'$. Parallel transport along $l(x,x')$ 
allows to define a canonical isomorphism between the tangent space 
$T_{x}\CD$ and $T_{x'}\CD$, which maps any given vector 
$v(x)\in T_{x}\CD$ into a corresponding vector 
$v_{P_{l(x,x')}}\in T_{x'}\CD$. If $\{\zeta _{(h)}(x) \}_{h=1,2,3}\in T_{x}\CD$ 
and $\{\zeta_{(k')}(x') \}_{k'=1,2,3}\in T_{x'}\CD$, respectively, denote   basis vectors, then the components of 
$v_{P_{l(x,x')}}$ can be expressed as 
\begin{equation}
\left(v_{P_{l(x,x')}}\right)^{k'}(x')=\tau ^{k'}_{h}(x',x)\,v^{h}(x)\;,
\end{equation}
where $\tau ^{k'}_{h}(x',x)$ denotes the bitensor 
$\in T_{x'}{\cal D}\otimes T_{x}^{*}{\cal D} $ associated with the parallel transport 
along $l(x,x')$. This characterizes the Dirac bitensorial measure in 
$U\subset \cal D $ according to
\begin{equation}
\delta ^{k'}_{h}(x',x):= \tau ^{k'}_{h}(x',x)\,\delta (x',x)\; ,
\end{equation}
where $\delta (x',x)$ is the standard Dirac distribution over the Riemannian manifold 
$({\cal D} ,g)$ (see \cite{lichnerowicz}). 
The Dirac bitensor $\delta ^{k'}_{h}(x',x)$ so defined is the elementary disturbance 
generating the Green function for the vector Laplacian  
$\Delta ^{a}_{\;b}$ according to 
\begin{equation}
_{x}\Delta^{a}_{\;h}\;E^{k'}_{a}(x',x)=\,-\,\delta _{h}^{k'}\,(x',x)\;,
\label{mindistgreen}
\end{equation}
where the pedex $x$ denotes the variable on which the vector Laplacian is acting. In terms of $E^{k}_{a}(p,q)$ we can write 
\begin{equation}
\psi ^{(i)}_{a}(x)=\,\lambda _{1}\,\int_{{\Sigma }_{x'}  }\,E^{k'}_{a}(x',x)\; \psi^{(i)}_{k'}(x')\,d\mu _{g}(x'),
\label{WGreenH}
\end{equation}
which, since $\int_{\Sigma}\psi ^{(i)}_{a}(x)\,\psi _{(k)}^{a}(x)\,d\mu _{g}=|\Sigma|\,\delta ^{i}_{k}$, yields
\begin{equation}
\frac{\lambda_{1}}{|\Sigma|}\,\int_{{\Sigma }_{x}  }\,\int_{{\Sigma }_{x'}  }\;{{\psi }}^{(k)}_{h'}(x')\,\,E^{h'}_{a}(x',x)\;
{{\psi}}_{(i)}^{a}(x)\,d\mu _{g}(x')\,d\mu _{g}(x)\,=\delta ^{k}_{i}\;.
\end{equation}
This latter remark implies that
 
\begin{equation}
\fl
p^{(k)}_{(i)}(x',x)\,d\mu _{g}(x')\,d\mu _{g}(x):=
\,\frac{\lambda_{1}}{|\Sigma|}\;{{\psi }}^{(k)}_{h'}(x')\,\,E^{h'}_{a}(x',x)\;
{{\psi}}_{(i)}^{a}(x)\,d\mu _{g}(x')\,d\mu _{g}(x)
\;,
\label{prob1}
\end{equation}
is a bivariate probability density describing the distribution of the geometric shear--anysotropies in $({\CD} ,g)$, along the minimal shear frame $x\mapsto \{\vec{\psi}_{(h)}\}_{h=1,2,3}$. If we define the components of the Green function with respect to  $\{\vec{\psi}_{(h)}\}_{h=1,2,3}$ according to
\begin{equation}
{\mathbb E}^{k}_{i}(x',x):=\,{{\psi }}^{(k)}_{h'}(x')\,\,E^{h'}_{a}(x',x)\;
{{\psi}}_{(i)}^{a}(x)\;,
\end{equation}
(the \emph{minimal shear Green function}), then we can rewrite (\ref{prob1}) in the simpler form
\begin{equation}
\fl
p^{(k)}_{(i)}(x',x)\,d\mu _{g}(x')\,d\mu _{g}(x):=
\,\frac{\lambda_{1}}{|\Sigma|}\;\,{\mathbb E}^{k}_{i}(x',x)\;\,d\mu _{g}(x')\,d\mu _{g}(x)
\;.
\label{prob2}
\end{equation}

\medskip

\noindent It is with respect to the distribution of geometric shear described by $p^{(k)}_{(i)}(x',x)\,d\mu _{g}(x')\,d\mu _{g}(x)$ that we can measure covariances and correlations of the shear generated, according to (\ref{vectW}),  by the distribution of currents  $\frac{2}{3}\nabla_{k}\theta- 8\pi GJ_{k}$. To this end let us consider the expression obtained from $\nabla ^{a}\sigma _{\parallel ab} $ 
by contracting with the solution $w^{b}$ of (\ref{vectW}) and integrating over $(\Sigma  ,g_{ab})$, \emph{i.e.},
\begin{equation}
\int_{\Sigma }w^{b}\nabla ^{a}(\sigma _{\parallel ab})\,d\mu _{g}
=\frac{2}{3}\int_{\Sigma   }w^{b}\nabla _{b}\theta \,d\mu _{g}-8\pi G
\int_{\Sigma   }w^{b}J_{b}\,d\mu _{g}\;. 
\label{intW}
\end{equation} 
Integrating by parts, and  exploiting again  the identity $ \sigma _{\parallel\, ab}\,\nabla ^{a}
w^{b}$ $=$ $\frac{1}{2}\,\sigma _{\parallel\, ab}\,\sigma _{\parallel}^{\,ab}$,  we easily get
\begin{equation}
\int_{\Sigma  }\sigma _{\parallel }{}^{2}\,\,d\mu _{g}=  
8\pi G\int_{\Sigma   }w^{b}J_{b}\,d\mu _{g}-\frac{2}{3}\int_{\Sigma   } w^{b}\,\nabla
_{b}\theta \,d\mu _{g}  \;,  
\label{intRel}
\end{equation}
which, by taking the average with respect to $(\Sigma,g)$, provides
\begin{equation}
\langle\sigma _{\parallel }{}^{2}\rangle_{\Sigma  }=\,8\pi G\langle w^{b}J_{b}\rangle_{\Sigma  } 
-\frac{2}{3}\langle w^{b}\,\nabla
_{b}\theta\rangle_{\Sigma  }\;.  
\label{intRelAv}
\end{equation}
In order to factorize the various contributions to $\langle\sigma _{\parallel }{}^{2}\rangle_{\Sigma  }$ coming from the various components of the current $8\pi G\,J_{b}-\frac{2}{3}\,\nabla_{b}\theta$,
let us consider  the integral representation of the solution of (\ref{vectW}) along the minimal shear Green function 
(\ref{mindistgreen}), \emph{i.e.},
\begin{equation}
w_{a}(x)=\,-\,\int_{{\Sigma }_{x'}  }\,{\mathbb E}^{k}_{a}(x',x)\;{\tilde{j}}_{k}(x')\,d\mu _{g}(x')
\label{WGreen}
\end{equation}
where, for notational ease, we have introduced the components ${\tilde{j}}_{k}(x')$ of the shear--generating current density with respect to the eigen--vectors basis $\{\vec{\psi}_{(k)}\}$, according to
\begin{equation}
{\tilde{j}}_{k}(x'):= \psi_{(k)}^{h'}(x')\, \left(\frac{2}{3}\nabla
_{h'}\theta- 8\pi GJ_{h'}  \right)_{x'}.
\label{sheargenerator}
\end{equation}

\begin{figure}[h]
\begin{center}
\includegraphics[bb= 0 0 540 470,scale=.5]{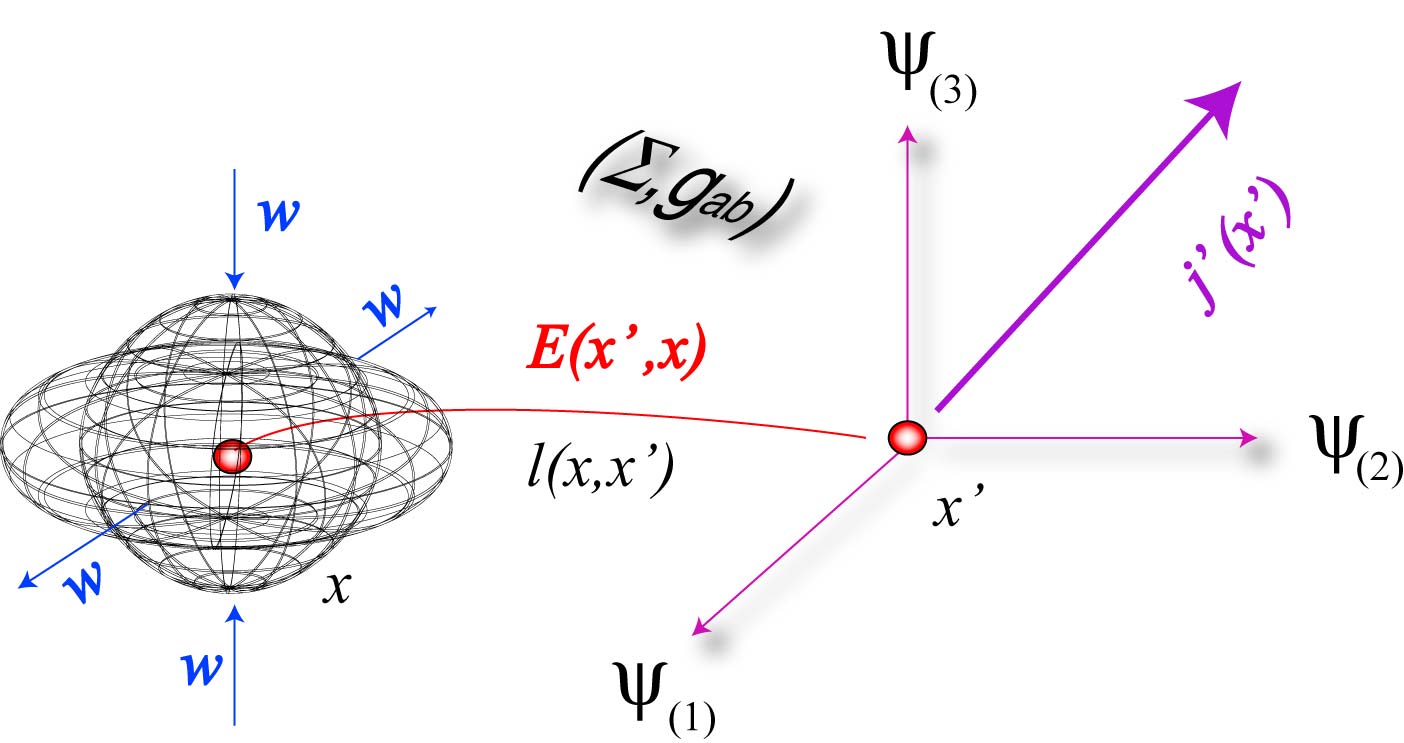}
\caption{The ``current" ${\tilde{j}}_{k}(x'):=\psi_{(k)}^{h'}(x')\, \left(\frac{2}{3}\nabla
_{h'}\theta- 8\pi GJ_{h'}  \right)_{x'}$, at the point $x'$, generates a shearing vector field $\vec{w}(x)$, at the point $x$, by means of the Green function ${\mathbb E}^{k}_{a}(x',x)$ of the vector Laplacian $\Delta ^{a}_{h}$. In this illustration, the points $x$ and $x'$ are separated by a geodesic segment $l(x,x')$, and the shearing vector field $\vec{w}$ is shown as deforming a spherical domain, centered on $x$, into an ellipsoid.}
\end{center}
\end{figure}

If we introduce (\ref{WGreen}) into (\ref{intRel}) we get
\begin{equation}
\int_{\Sigma  }\sigma _{\parallel }{}^{2}\,d\mu _{g}=
\int_{{\Sigma }_{x}  }\,\int_{{\Sigma }_{x'}  }\;{\tilde{j}}_{k}(x')\,\,{\mathbb E}^{k}_{a}(x',x)\;{\tilde{j}}^{a}(x)\,d\mu _{g}(x')\,d\mu _{g}(x)\;,
\end{equation}
which, according to (\ref{prob2}), implies 
\begin{equation}
\fl\quad
\int_{\Sigma   }\sigma _{\parallel }{}^{2}\,d\mu _{g}=\,
\frac{|\Sigma |}{\lambda_{1}}\,\,\int_{{\Sigma }_{x}  }\,\int_{{\Sigma }_{x'}  }\;
{\tilde{j}}_{b}(x')\;p^{(b)}_{(a)}(x',x)\;{\tilde{j}}^{a}(x)\,d\mu _{g}(x')\,d\mu _{g}(x)\;.
\label{intrepbella}
\end{equation}
Note that the expression
\begin{equation}
\fl\quad
Cov\left({\tilde{j}},\,{\tilde{j}}\right):=\,
\int_{{\Sigma }_{x}  }\,\int_{{\Sigma }_{x'}  }\;
{\tilde{j}}_{b}(x')\;p^{(b)}_{(a)}(x',x)\;{\tilde{j}}^{a}(x)\,d\mu _{g}(x')\,d\mu _{g}(x)
\label{covjj}
\end{equation}
can be naturally interpreted 
as the covariance, in the minimal shear frame $({\CD },\{\vec{\psi}_{(i)}\})$, between 
the values of the current ${\tilde{j}}$ attained at the point $x\in \cal D $ versus the values 
attained in neighbouring points $y\not=x$, as these points vary over $\cal D $. 
Thus (\ref{intrepbella}) can be rewritten as 
\begin{equation}
\int_{\Sigma  }\sigma _{\parallel }{}^{2}\,d\mu _{g}=
\;\frac{|\Sigma |}{\lambda_{1}}\;Cov\left({\tilde{j}},\,{\tilde{j}}\right)\;,
\end{equation}
or, more explicitly, in terms of the factors $J$ and $\nabla \theta $, (see (\ref{sheargenerator})), 
\begin{equation}
\fl
\int_{\Sigma  }\sigma _{\parallel }{}^{2}\,d\mu _{g}=
=\,\frac{|\Sigma |}{\lambda_{1}}\;\left[\frac{4}{9}Cov(\nabla \theta,\nabla \theta)+ (8\pi\, G)^{2}\,Cov(J,J)
-\frac{32\pi \,G}{3}Cov(\nabla \theta, \,J) \right]\;,\nonumber
\end{equation}
where $Cov(\circ ,\bullet  )$ is defined from (\ref{covjj}) in an obvious way.

\begin{figure}[h]
\begin{center}
\includegraphics[bb= 0 0 540 470,scale=.5]{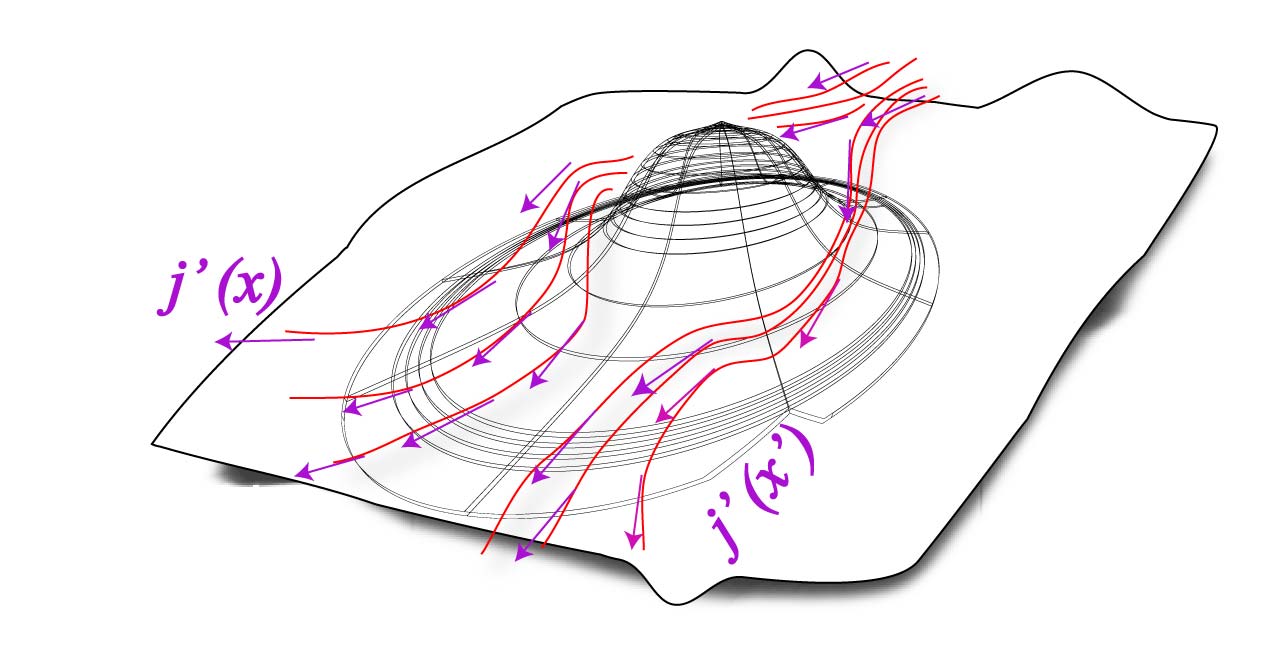}
\caption{The Green function ${\mathbb E}^{k}_{a}(x',x)$ of the vector Laplacian $\Delta ^{a}_{h}$ can be used to describe the curvature induced correlations among the distribution of the shear generating ``currents" ${\tilde{j}}_{k}(x')$ and ${\tilde{j}}_{k}(x)$. By averaging these correlations over the domain of near--homogeneity one obtains the various correlation lengths, which describe the different contributions to the averaged longitudinal shear.}
\end{center}
\end{figure}

As customary, we standardize (the absolute values of) these covariances between $0$ and $1$ by dividing them by the appropriate powers of 
$\left\langle \left|{{J}}\right| ^{2}\right\rangle _{\cal D }$ and $\left\langle \left|{{\nabla\theta}}\right| ^{2}\right\rangle _{\cal D }$, (we are here tacitly assuming that the main contribution to the shear generating  currents comes from $\CD $, \emph{i.e.}, $\left\langle \left|{{J}}\right| ^{2}\right\rangle _{\Sigma }=\left\langle \left|{{J}}\right| ^{2}\right\rangle _{\CD }$, and $\left\langle \left|{{\nabla\theta}}\right| ^{2}\right\rangle _{\Sigma}=\left\langle \left|{{\nabla\theta}}\right| ^{2}\right\rangle _{\cal D }$). This defines the corresponding correlation coefficients
$$
Corr\left({{\nabla\theta}},\,{{\nabla\theta}}\right):=
\frac{Cov\left({{\nabla\theta}},\,{{\nabla\theta}}\right)}{\left\langle \left|{{\nabla\theta}}\right| ^{2}\right\rangle _{\cal D }}\;\;;\;\;
Corr\left({{J}},\,{{J}}\right):=
\frac{Cov\left({{J}},\,{{J}}\right)}{\left\langle \left|{{J}}\right| ^{2}\right\rangle _{\cal D }}\;
$$
and
\begin{equation}
Corr\left({{\nabla\theta}},\,{{J}}\right):=
\frac{Cov\left({{\nabla\theta}},\,{{J}}\right)}{\left\langle \left|{{\nabla\theta}}\right| ^{2}\right\rangle _{\cal D }^{\frac{1}{2}}\,\left\langle \left|{{J}}\right| ^{2}\right\rangle _{\cal D }^{\frac{1}{2}}}\;.
\end{equation}
Note that $0\leq Corr\left({{\nabla\theta}},\,{{\nabla\theta}}\right)\leq 1$,  $0\leq Corr\left({{J}},\,{{J}}\right)\leq 1$, whereas 
$-1\leq Corr\left({{\nabla\theta}},\,{{J}}\right)\leq 1$. The scale of $\CD $ is set by $\lambda_{1}^{-\frac{1}{2}}$, (see (\ref{condscale}), thus it follows that to such correlations we can associate the corresponding \emph{correlation lengths} according to
$$
L_{\nabla \theta ,\nabla \theta }:=\,\left[\frac{Corr\left({{\nabla\theta}},\,{{\nabla\theta}}\right)}{\lambda_{1}}\right]^{\frac{1}{2}}\;\;;\;\;
L_{J ,J}:=\,\left[\frac{Corr\left({{J}},\,{{J}}\right)}{\lambda_{1}}\right]^{\frac{1}{2}}\;,
$$
and
\begin{equation}
L_{\nabla \theta ,J }:=\,\left|\frac{Corr\left({{\nabla\theta}},\,{{J}}\right)}{\lambda_{1}}\right|^{\frac{1}{2}}\;,
\end{equation}
which provide the {\em length scales} over which the distributions of  ${\nabla\theta }$, and $J$ are significantly correlated and thus  contribute appreciably to the {\emph {matter--expansion shear}} $\sigma _{\parallel }{}^{2}$ in ${\cal D}$ (see Figure 15).

\begin{figure}[h]
\begin{center}
\includegraphics[bb= 0 0 540 470,scale=.5]{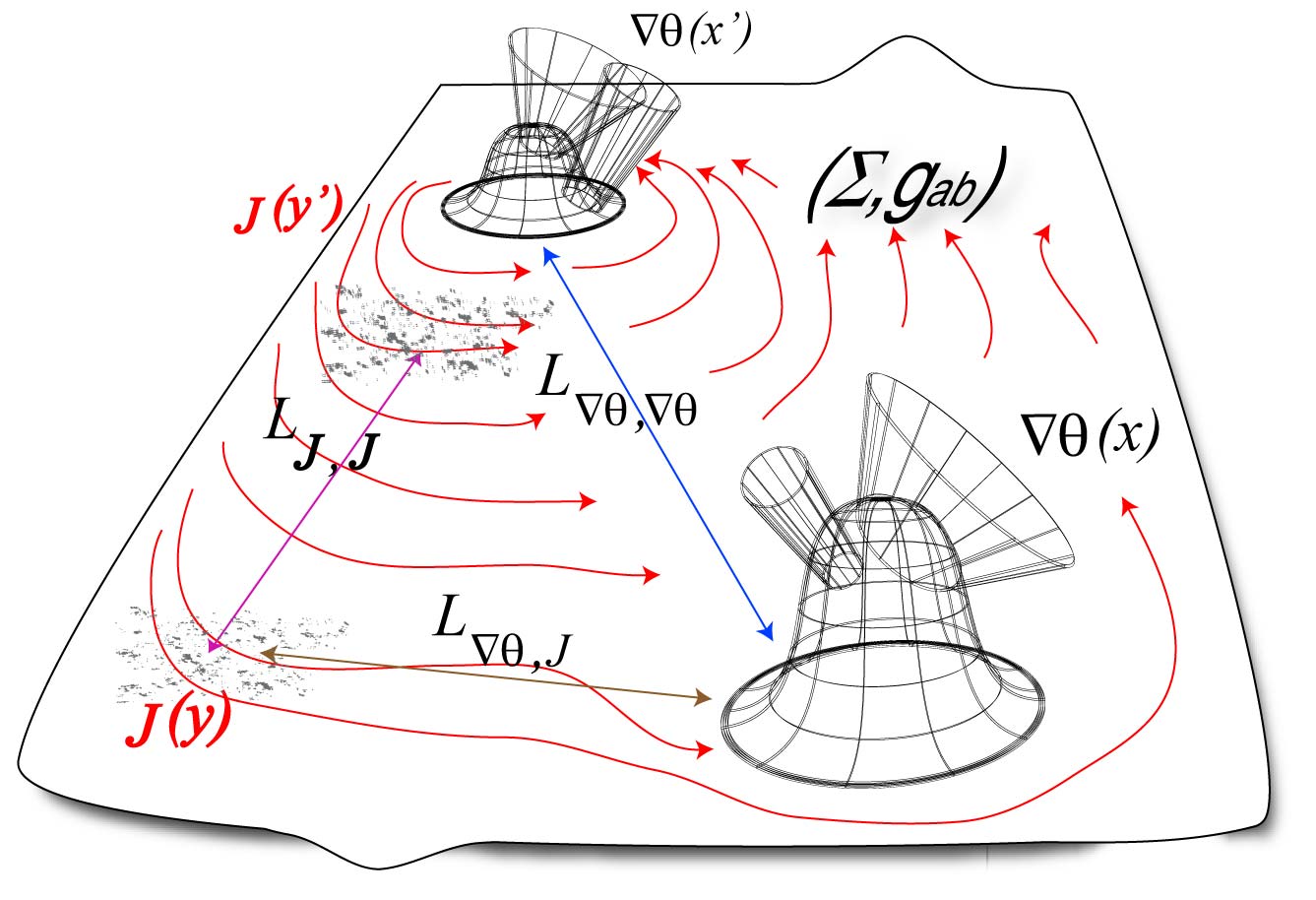}
\caption{A pictorial rendering of the various correlation lengths associated with the local distribution of the matter current $\vec{J}$ and the distribution of the gradient of the rate of expansion $\nabla \theta$.}
\end{center}
\end{figure}

In terms of these correlation lengths we can  write, (dividing both members by $|\Sigma |$ and assuming that $\left\langle \sigma _{\parallel}{} ^{2}\right\rangle _{\Sigma}=\left\langle \sigma _{\parallel}{} ^{2}\right\rangle _{\CD }$),

\begin{eqnarray}
\left\langle \sigma _{\parallel}{} ^{2}\right\rangle _{\CD }=\,\frac{4}{9}L^{2}_{\nabla \theta,\nabla \theta}\left\langle \left|{{\nabla\theta}}\right| ^{2}\right\rangle _{\cal D }+ (8\pi\, G)^{2}\,L^{2}_{J,J}\left\langle \left|{{J}}\right| ^{2}\right\rangle _{\cal D } \label{mcd}\\
\nonumber\\
\mp \;\frac{32\pi \,G}{3}L^{2}_{\nabla \theta, \,J}\left\langle \left|{{\nabla\theta}}\right| ^{2}\right\rangle _{\cal D }^{\frac{1}{2}}\,\left\langle \left|{{J}}\right| ^{2}\right\rangle _{\cal D }^{\frac{1}{2}}\;,\nonumber
\end{eqnarray}

\medskip

\noindent where $\mp $ is the sign of $-\,Corr\left({{\nabla\theta}},\,{{J}}\right)$.
 To extract useful information from this relation let us remark that
\begin{equation}
L_{\delta H_{\CD}}:= \,
\left( \frac{\int_{\cal D  }\left(\theta -\left\langle  \,
\theta \right\rangle _{\cal D }\right)^{2}\,\,d\mu _{g}}{\int_{\cal D}|\nabla \theta|{}^{2} \,\,d\mu _{g}} \right)^{\frac{1}{2}}\;,
\end{equation}
is the typical  length scale associated with the spatial fluctuations of the expansion $\frac{1}{3}\theta $ 
with respect to the average Hubble parameter in $\CD$. Since $\left(\left\langle\theta ^{2}\right\rangle _{\cal D }-\left\langle\theta \right\rangle _{\cal D }^{2}\right)=
9\,\delta ^{2}H_{\cal D}$, we can write
\begin{equation}
\left\langle \left| \nabla \,
\theta \right| ^{2}\right\rangle _{\cal D }=9\,L^{-\,2}_{\delta H_{\CD}}\;\delta ^{2}H_\CD\;.
\label{flucthetaD}
\end{equation}
If necessary, this expression for $\left\langle \left| \nabla \,
\theta \right| ^{2}\right\rangle _{\cal D }$ can be resolved into its finer components in the overdense and underdense domains into which $\CD$ is factorized.

\begin{figure}[h]
\begin{center}
\includegraphics[bb= 0 0 540 470,scale=.5]{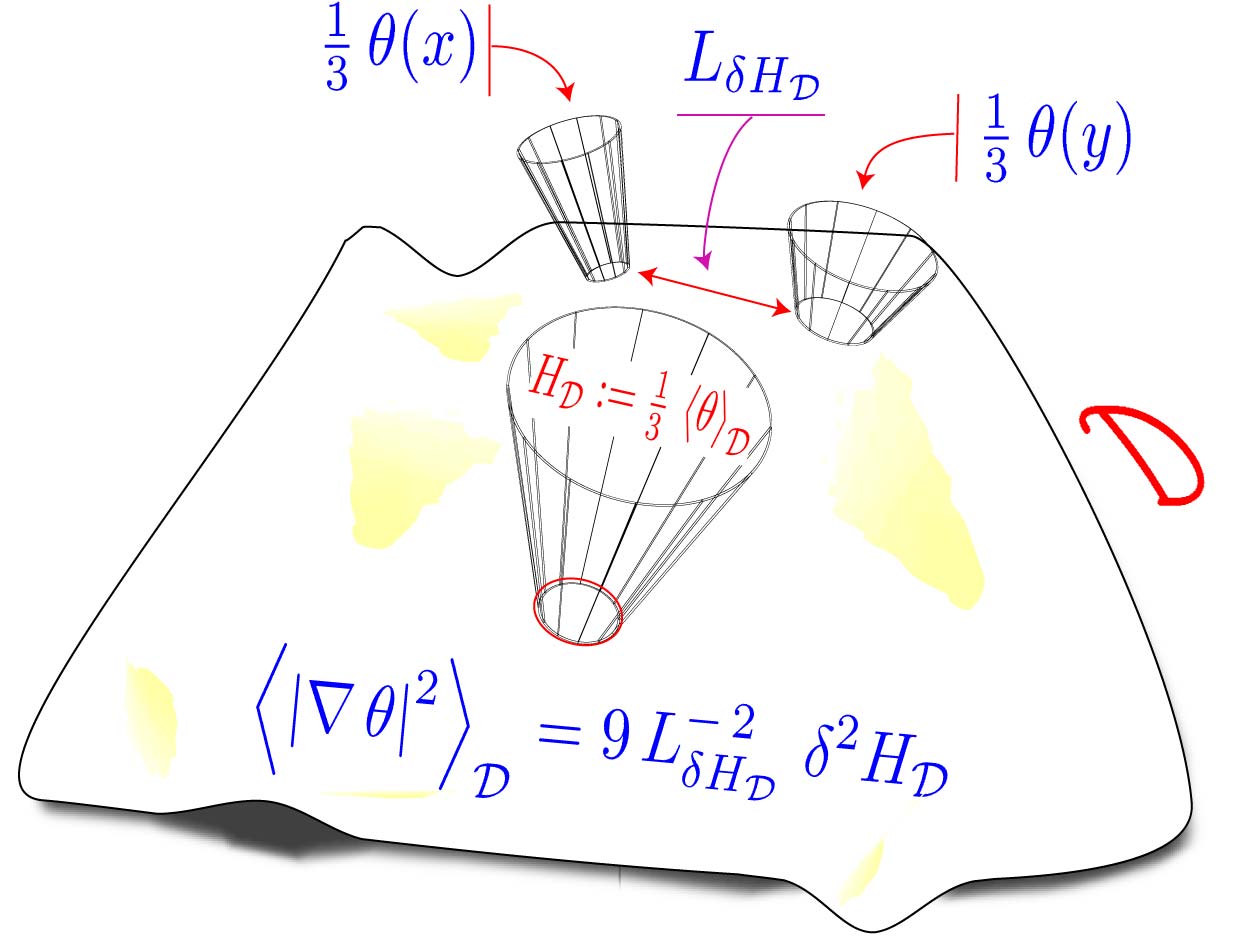}
\caption{A representation of the length scale $L_{\delta H_{\CM}}$ associated with the spatial fluctuations of the local expansion $\frac{1}{3}\theta $ with respect to the average Hubble parameter $H_{\cal D}$ in $\CD$, and its relation with the variance $\delta^{2}H_{\cal D}$ of $H_{\cal D}$ and the gradient of the expansion $\nabla \theta$. The resulting  expression can be factorized, (see (\ref{gradesp})), in its contributions coming from the underdense regions $\cal E$ and the matter--dominated regions $\cal M$ partitioning the domain of near--homogeneity $\cal D$.}
\end{center}
\end{figure}

Let us define 
\begin{equation}
\fl
L_{\delta H_{\CM}}:= \,
\left( \frac{\int_{\cal M  }\left(\theta -\left\langle  \,
\theta \right\rangle _{\cal M }\right)^{2}\,\,d\mu _{g}}{\int_{\cal M}|\nabla \theta|{}^{2} \,\,d\mu _{g}} \right)^{\frac{1}{2}}\;,\;\;\;L_{\delta H_{\CE}}:= \,
\left( \frac{\int_{\cal E }\left(\theta -\left\langle  \,
\theta \right\rangle _{\cal E}\right)^{2}\,\,d\mu _{g}}{\int_{\cal E}|\nabla \theta|{}^{2} \,\,d\mu _{g}} \right)^{\frac{1}{2}}\;,
\end{equation}
which, as above, can be interpreted as the typical  length scales associated with the spatial fluctuations of the expansion $\frac{1}{3}\theta $ 
with respect to the local averages of the Hubble parameters in the overdense and underdense domains ${\cal M}$ and $\CE$ (see Figure 16). Thus, if we  factorize $\left\langle \left| \nabla \,
\theta \right| ^{2}\right\rangle _{\cal D }$ as   
\begin{eqnarray}
\fl
\left\langle \left| \nabla \,
\theta \right| ^{2}\right\rangle _{\cal D }=\lambda_{\CM }\,\frac{\left\langle \left| \nabla \,
\theta \right| ^{2}\right\rangle _{\cal M }}{\left\langle  \,
\theta ^{2}\right\rangle _{\cal M}-\left\langle  \,
\theta \right\rangle _{\cal M }^{2}}\;\left(\left\langle  \,
\theta ^{2}\right\rangle _{\cal M }-\left\langle  \,
\theta \right\rangle _{\cal M}^{2}\right) \nonumber\\
+\;
(1-\lambda_{\CM })\,\frac{\left\langle \left| \nabla \,
\theta \right| ^{2}\right\rangle _{\cal E }}{\left\langle  \,
\theta ^{2}\right\rangle _{\cal E}-\left\langle  \,
\theta \right\rangle _{\cal E }^{2}}\;\left(\left\langle  \,
\theta ^{2}\right\rangle _{\cal E }-\left\langle  \,
\theta \right\rangle _{\cal E }^{2}\right)\;,
\end{eqnarray}
we can express the term $\left\langle \left|\nabla \theta \right| ^{2}\right\rangle _{\cal D }$ as

\begin{equation}
\fl
\left\langle \left| \nabla \,
\theta \right| ^{2}\right\rangle _{\cal D }=\lambda_{\CM }\,\,\,L^{-\,2}_{\delta H_{\CM}}\;\,\left(\left\langle  \,
\theta ^{2}\right\rangle _{\cal M }-\left\langle  \,
\theta \right\rangle _{\cal M }^{2}\right)+
(1-\lambda_{\CM })\,L^{-\,2}_{\delta H_{\CE}}\;\,\left(\left\langle  \,
\theta ^{2}\right\rangle _{\cal E }-\left\langle  \,
\theta \right\rangle _{\cal E }^{2}\right)
\;.
\end{equation}
Since $\left(\left\langle\theta ^{2}\right\rangle _{\cal M }-\left\langle\theta \right\rangle _{\cal M }^{2}\right)=
9\,\delta ^{2}H_{\cal M}$ and $\left(\left\langle\theta ^{2}\right\rangle _{\cal E }-\left\langle\theta \right\rangle _{\cal E }^{2}\right)=
9\,\delta ^{2}H_{\cal E}$ we can re--express (\ref{flucthetaD}) as

\begin{equation}
\fl
\left\langle \left| \nabla \,
\theta \right| ^{2}\right\rangle _{\cal D }=9\,L^{-\,2}_{\delta H_{\CD}}\;\delta ^{2}H_{\CD}=9\,\lambda_{\CM }\,L^{-\,2}_{\delta H_{\CM}}\;\delta ^{2}H_\CM +9\,(1-\lambda_{\CM })\,\,L^{-\,2}_{\delta H_{\CE}}\;\delta ^{2}H_\CE \;,
\label{gradesp}
\end{equation}

\noindent which implies the useful relation

\begin{equation}
\delta ^{2}H_{\CD}=\,\lambda_{\CM }\,\frac{L^{\,2}_{\delta H_{\CD}}}{L^{\,2}_{\delta H_{\CM}}}\,\delta ^{2}H_{\CM}+
\,(1-\lambda_{\CM })\,\,\frac{L^{\,2}_{\delta H_{\CD}}}{L^{\,2}_{\delta H_{\CE}}}\,\delta ^{2}H_{\CE}\;.
\end{equation}

In order to deal similarly with the matter--current term $\left\langle \left|J\right| ^{2}\right\rangle _{\cal D }$ appearing in (\ref{mcd}), it is profitable to rewrite it as 
\begin{equation}
\left\langle \left|J\right| ^{2}\right\rangle _{\cal D }=\lambda _{\CM}\,\frac{\left\langle \left|J\right| ^{2}\right\rangle _{\cal M }}{\left\langle \varrho \right\rangle^{2} _{\cal M }} \,\left\langle \varrho \right\rangle^{2} _{\cal M }
+(1-\lambda _{\CM})\,\frac{\left\langle \left|J\right| ^{2}\right\rangle _{\cal E }}{\left\langle \varrho \right\rangle^{2} _{\cal E }} \,\left\langle \varrho \right\rangle^{2} _{\cal E}
\;,
\end{equation}
where $\left\langle \varrho \right\rangle _{\cal M }$ and  $\left\langle \varrho \right\rangle _{\cal E }$ respectively denote the mass density in the overdense and underdense regions  $\cal M$ and $\cal E$. Let us introduce the adimensional ratios, (we are using units in which $c=1$),
\begin{equation*}
V^{2}_{\varrho }[\CM]:= 
\frac{\left\langle \left|J\right| ^{2}\right\rangle _{\cal M }}{\left\langle \varrho \right\rangle^{2} _{\cal M }}=
\frac{\sum_{i}\,v^{2}_{\CM^{(i)}}\left\langle \varrho \right\rangle^{2} _{{\cal M}^{(i)}}\,\left|\CM^{(i)} \right|}{\sum_{k}\,\left\langle \varrho \right\rangle^{2} _{{\cal M}^{(k)}}\,\left|\CM^{(k)} \right|}
\;;
\end{equation*}
\begin{equation}
V^{2}_{\varrho }[\CE]:= 
\frac{\left\langle \left|J\right| ^{2}\right\rangle _{\cal E }}{\left\langle \varrho \right\rangle^{2} _{\cal E }}=
\frac{\sum_{\alpha }\,v^{2}_{\CE^{(\alpha )}}\left\langle \varrho \right\rangle^{2} _{{\cal E}^{(\alpha )}}\,\left|\CE^{(\alpha )} \right|}{\sum_{\beta }\,\left\langle \varrho \right\rangle^{2} _{{\cal E}^{(\beta )}}\,\left|\CE^{(\beta )} \right|}
\;, 
\end{equation}
with $v^{2}_{\CM^{(i)}}:=\frac{\left\langle \left|J\right| ^{2}\right\rangle _{\CM^{(i)}}}{\left\langle \varrho \right\rangle^{2} _{{\cal M}^{(i)}}}$, ($:=0$ if $\left\langle \varrho \right\rangle^{2} _{{\cal M}^{(i)}}:= 0$), and
$v^{2}_{\CE^{(\alpha )}}:=\frac{\left\langle \left|J\right| ^{2}\right\rangle _{\CE^{(\alpha )}}}{\left\langle \varrho \right\rangle^{2} _{{\cal E}^{(\alpha )}}}$, ($:=0$ if $\left\langle \varrho \right\rangle^{2} _{{\cal E}^{(\alpha )}}:= 0$). Since $V^{2}_{\varrho }[\CM]$ and $V^{2}_{\varrho }[\CE]$ are weighted averages of the (squared norms of the) typical velocity of matter in the overdense and underdense regions $\CM^{(i)}$ and $\CE^{(\alpha )}$, we 
can naturally interpret $V^{2}_{\varrho }[\CM]$ and $V^{2}_{\varrho }[\CE]$ as the (squared norm of the) typical velocity of matter in the respective  matter--dominated portion $\cal M$ and vacuum--dominated portion $\cal E$ of $\CD$ (see Figure 17). Thus, we write
\begin{equation}
\left\langle \left|J\right| ^{2}\right\rangle _{\cal D }=\lambda _{\CM}\,V^{2}_{\varrho }[\CM]\,\left\langle \varrho \right\rangle^{2} _{\cal M }+(1-\lambda _{\CM})\,V^{2}_{\varrho }[\CE]\,\left\langle \varrho \right\rangle^{2} _{\cal E }\;,
\end{equation}
with $V^{2}_{\varrho }[\CM]\leq 1$,\,$V^{2}_{\varrho }[\CE]\leq 1$. 
If we normalize this latter expression by $\left\langle \varrho \right\rangle^{2} _{\cal D }$ then we can  define the (squared) typical velocity of matter in the region $\CD$ according to
\begin{equation}
V^{2}_{\varrho }[\CD]\,:= \frac{\left\langle \left|J\right| ^{2}\right\rangle _{\cal D }}{\left\langle \varrho \right\rangle^{2} _{\cal D }}=\,\lambda _{\CM}\,V^{2}_{\varrho }[\CM]\,\frac{\left\langle \varrho \right\rangle^{2} _{\cal M }}{\left\langle \varrho \right\rangle^{2} _{\cal D }}+(1-\lambda _{\CM})\,V^{2}_{\varrho }[\CE]\,\frac{\left\langle \varrho \right\rangle^{2} _{\cal E }}{\left\langle \varrho \right\rangle^{2} _{\cal D }}\;.
\end{equation}

\begin{figure}[h]
\begin{center}
\includegraphics[bb= 0 0 540 470,scale=.4]{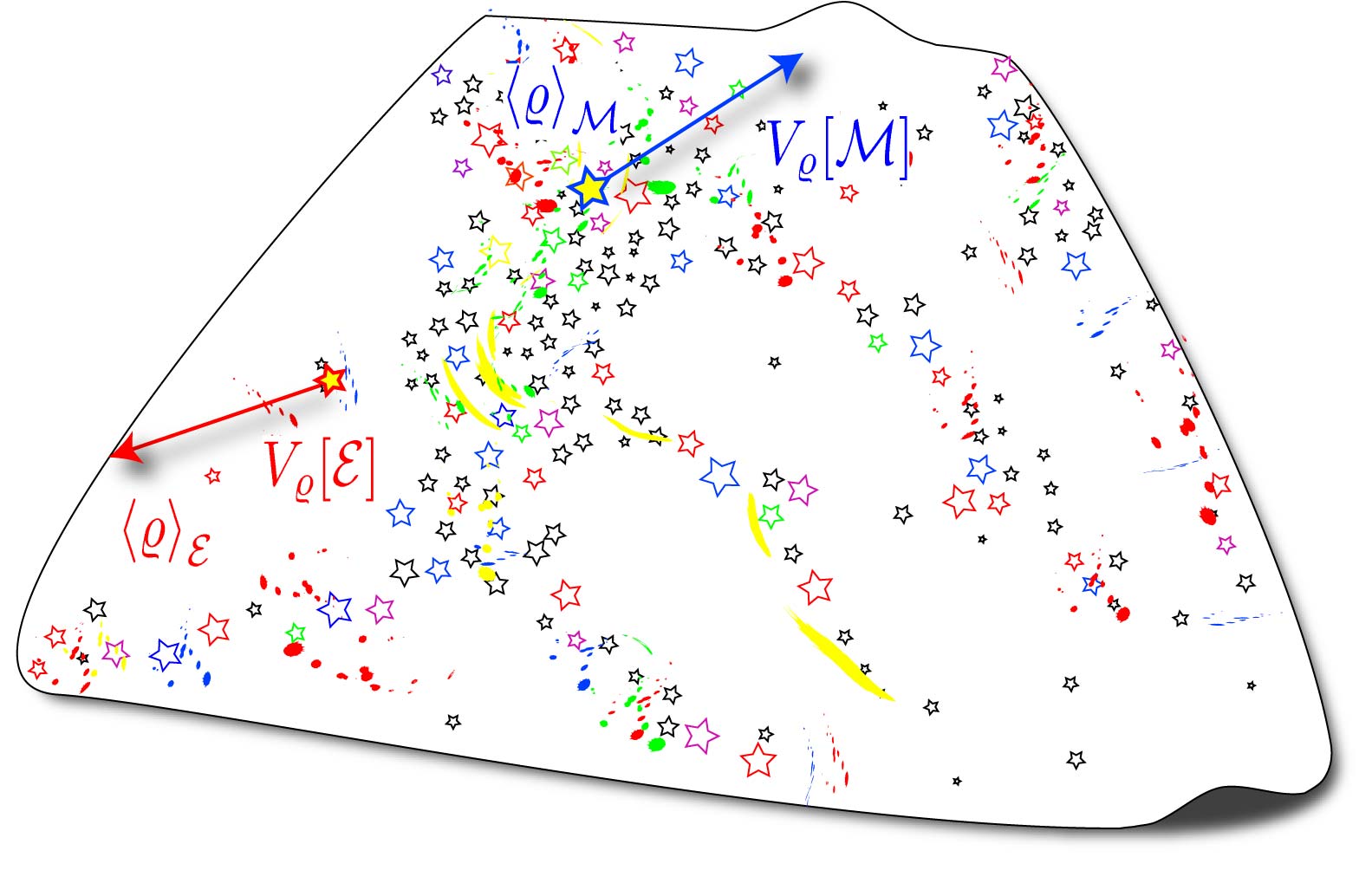}
\caption{The average velocities $V_{\varrho }[\CM]$ and $V_{\varrho }[\CE]$ in the matter--dominated $\CM$ and in the underdense region $\CE$, respectively. These quantities are defined in terms of the corresponding average densities
$\left\langle \varrho \right\rangle _{\cal M }$ and $\left\langle \varrho \right\rangle _{\cal E }$.}
\end{center}
\end{figure}

\medskip

\noindent Inserting these parametrizations  into (\ref{mcd}), and normalizing to the squared effective Hubble parameter $H_{\cal D }^{\,2}$, we eventually get

\begin{eqnarray}
\fl
\frac{\langle \sigma _{\parallel }{}^{2}\rangle_{\cal D }}{H_{\cal D }^{\,2}}=\,
\lambda _{\CM}\,\left[4\,\frac{L_{\nabla\theta,\nabla\theta}^{2}}{L_{\delta H_{\CM}}^{2}}\,\left(\frac{\delta ^{2}H_{\CM}}{H_{\cal D }^{\,2}}\right)+\,V^{2}_{\varrho }[\CM]\,
\frac{L_{J,J}^{2}}{L_{\delta H_{\CM}}^{2}}\,\left(    
\frac{(8\pi G)^{2}\,\left\langle \varrho \right\rangle^{2} _{\cal M }\,  L_{\delta H_{\CM}}^{2}}{H_{\cal D }^{\,2}}\right) \right] \label{MattShear}\\
\nonumber\\
+(1-\lambda _{\CM})\,
\left[4\,\frac{L_{\nabla\theta,\nabla\theta}^{2}}{L_{\delta H_{\CE}}^{2}}\,\left(\frac{\delta ^{2}H_{\CE}}{H_{\cal D }^{\,2}}\right)+\,V^{2}_{\varrho }[\CE]\,
\frac{L_{J,J}^{2}}{L_{\delta H_{\CE}}^{2}}\,\left(    
\frac{(8\pi G)^{2}\,\left\langle \varrho \right\rangle^{2} _{\cal E }\,  L_{\delta H_{\CE}}^{2}}{H_{\cal D }^{\,2}}\right) \right] \nonumber\\
\nonumber\\
\mp 
\,\,V^{2}_{\varrho }[\CD]\,\frac{L_{\nabla\theta,J}^{2}}{L_{\delta H_{\CD}}^{2}}\,
\left( \frac{32\pi G\,\left\langle \varrho \right\rangle _{\cal D } \,\;L_{\delta H_{\CD}}\,
\left(\delta ^{2}H_\CD\right)^{\frac{1}{2}}}{H_{\cal D }^{\,2}} \right) \nonumber\;.
\end{eqnarray}

\medskip
 
\subsection{The transverse gravitational shear}
 
\noindent  At this point, it is  important to stress that the norm of the transverse part $
2\sigma _{\perp }{}^{2}:= g^{ab}g^{cd}\sigma _{\perp ac}\sigma _{\perp bd}
$, is not determined by (\ref{vectW}). The term $\langle \sigma _{\perp }{}^{2}\rangle_{\cal D }$  is associated with the presence
of a non--trivial initial rate of variation of the conformal geometry of $({\cal D} ,g)$, 
\emph{i.e.}, with initial data describing the presence of
gravitational radiation of cosmological origin in $({\cal D} ,g)$ (see Figure 18). We define the total energy density of gravitational waves  in $\CD$ by
\begin{equation}
\varrho ^{\CD}_{GW}:= \frac{\langle \sigma _{\perp }{}^{2}\rangle_{\cal D }}{32\pi G}\;,
\quad{\rm so}\;\,{\rm that}\qquad
\frac{\langle \sigma _{\perp }{}^{2}\rangle_{\cal D }}{3\,H^{2}_{\CD}}=4\,\frac{\varrho ^{\CD}_{GW}}{\varrho^\CD_{crit}}\;,
\end{equation}
where $\varrho^\CD_{crit}:= \frac{3\,H^{2}_{\CD}}{8\pi G}$ is the density formally associated with the ``critical density'' of the standard Friedmannian model (in the region $\CD$ of near homogeneity). The ratio
\begin{equation}
 \frac{\varrho ^{\CD}_{GW}}{\varrho^\CD_{crit}}:= \Omega _{GW}^{\CD}\;,
\end{equation}
describing the relative strength of the energy density of gravitational waves with respect to the critical density, is the quantity conventionally used in cosmology for describing gravitational waves of cosmological origin. Thus, we parametrize the shear term $\langle \sigma _{\perp }{}^{2}\rangle_{\cal D }$ according to
\begin{equation}
\Omega _{GW}^{\CD} =
\frac{\langle \sigma _{\perp }{}^{2}\rangle_\CD}{12 H^{2}_{\CD}}\;.
\end{equation}

\begin{figure}[h]
\begin{center}
\includegraphics[bb= 0 0 540 470,scale=.4]{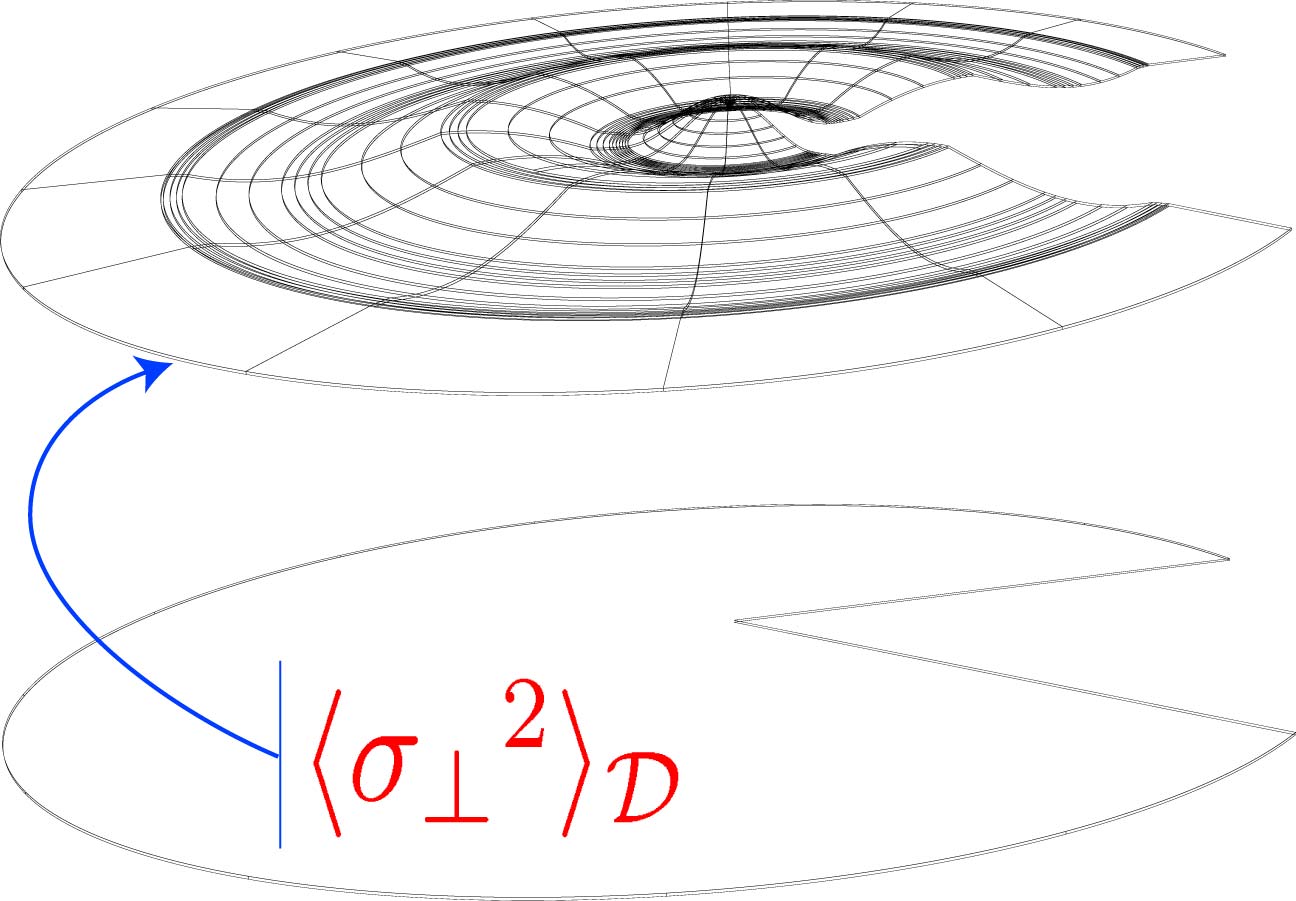}
\caption{The transverse shear $\langle \sigma _{\perp }{}^{2}\rangle_{\cal D }$ describes the rate of variation of the conformal geometry of $({\cal D} ,g)$, and can be associated with the presence of
gravitational radiation of cosmological origin in the region of near--homogeneity $({\cal D} ,g)$.}
\end{center}
\end{figure}

Equipped with the above estimates, we now consider the  backreaction term ${\cal Q}_\CD$, (divided 
by $6\,H_{{\cal D}}^{2}$). 
From (\ref{mexpl})
we  finally arrive at the key--result of this paper:

\begin{eqnarray}
\label{resultQ}
\frac{1}{6\,H_{{\cal D}}^{2}} \left( 
{\cal Q}_{\CD}+ 2 \langle \sigma _{\perp }{}^{2}\rangle_\CD\right)\;=\;-\Omega^\CD_\CQ +4\,\Omega _{GW}^{\CD} =
\nonumber\\
\fl
\lambda _{\CM}\,\left[\left(1-8\,\frac{L_{\nabla\theta,\nabla\theta}^{2}}{L_{\delta H_{\CM}}^{2}}\right)\,\left(\frac{\delta ^{2}H_{\CM}}{H_{\cal D }^{\,2}}\right)-2\,V^{2}_{\varrho }[\CM]\,
\frac{L_{J,J}^{2}}{L_{\delta H_{\CM}}^{2}}\,\left(    
\frac{(8\pi G)^{2}\,\left\langle \varrho \right\rangle^{2} _{\cal M }\,  L_{\delta H_{\CM}}^{2}}{H_{\cal D }^{\,2}}\right) \right] \nonumber\\
\nonumber\\
\fl
+(1-\lambda _{\CM})\,
\left[\left(1-8\,\frac{L_{\nabla\theta,\nabla\theta}^{2}}{L_{\delta H_{\CE}}^{2}}\right)\,\left(\frac{\delta ^{2}H_{\CE}}{H_{\cal D }^{\,2}}\right)-2\,V^{2}_{\varrho }[\CE]\,
\frac{L_{J,J}^{2}}{L_{\delta H_{\CE}}^{2}}\,\left(    
\frac{(8\pi G)^{2}\,\left\langle \varrho \right\rangle^{2} _{\cal E }\,  L_{\delta H_{\CE}}^{2}}{H_{\cal D }^{\,2}}\right) \right] \nonumber\\
\nonumber\\
\fl
+{\lambda_\CM}(1-\lambda_\CM)\,\frac{\left(H_\CE-H_\CM\right)^{2}}{H_{{\cal D}}^{2}}
\pm  
\,\,2V^{2}_{\varrho }[\CD]\,\frac{L_{\nabla\theta,J}^{2}}{L_{\delta H_{\CD}}^{2}}\,
\left( \frac{32\pi G\,\left\langle \varrho \right\rangle _{\cal D } \,\;L_{\delta H_{\CD}}\,
\left(\delta ^{2}H_\CD\right)^{\frac{1}{2}}}{H_{\cal D }^{\,2}} \right) \nonumber\;.\\
\end{eqnarray}
 
\noindent together with the formula for the averaged scalar curvature (see Figure 19):
\begin{equation}
\label{resultR}
\Omega_\CR^\CD + \Omega^\CD_{\Lambda} = 1 - \Omega^\CD_m - \Omega^\CD_\CQ  \quad {\rm with}\quad \Omega^\CD_\CQ \quad {\rm from}\; {\rm above}\;.
\end{equation}

\begin{figure}[h]
\begin{center}
\includegraphics[bb= 0 0 540 470,scale=.4]{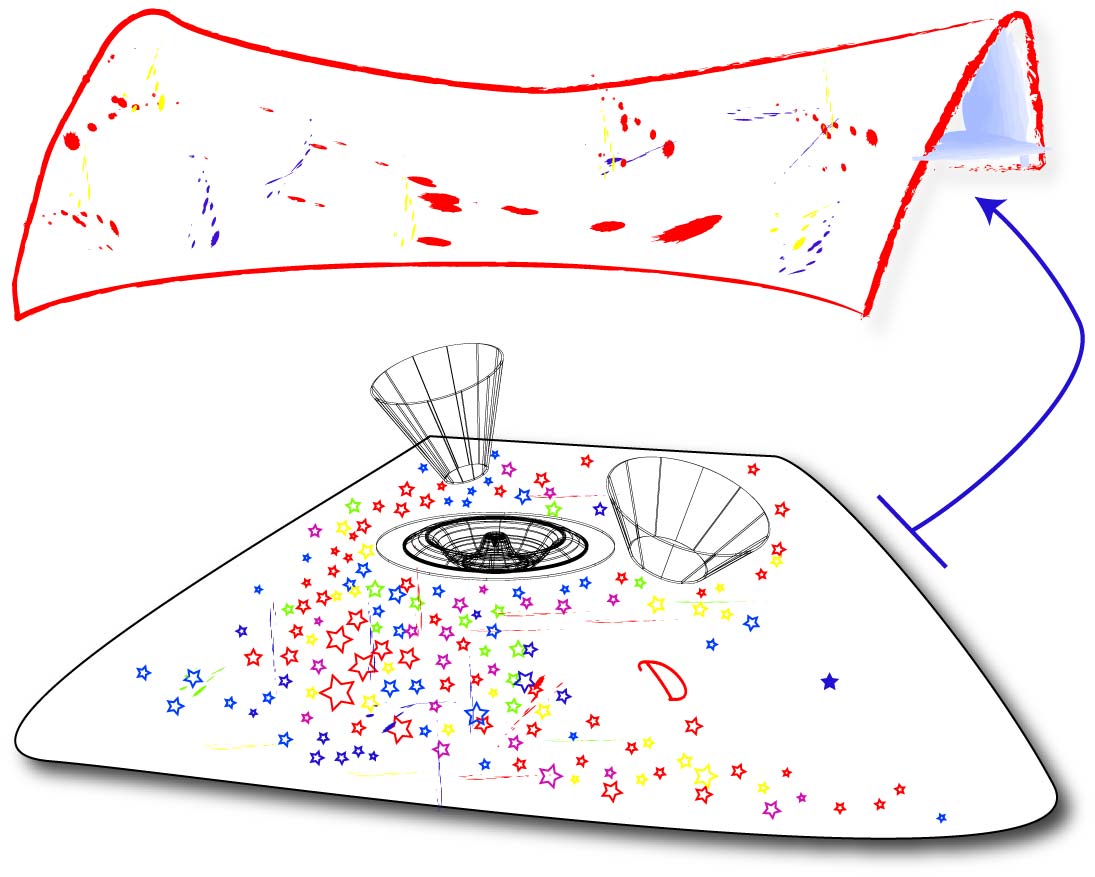}
\caption{The kinematical backreaction, as described by (\ref{resultQ}), has, in the course of evolution, led to the emergence of a dynamical curvature term $\Omega_\CR^\CD$ (\ref{resultR}) in the region of near--homogeneity $({\cal D} ,g)$.}  
\end{center}
\end{figure}

\subsection{Bound on kinematical backreaction in the matter--dominated Late Universe}
\label{subsection:boundsQ}

We are now going to invoke approximate assumptions in order to illustrate the result $\lbrace$(\ref{resultQ}, \ref{resultR})$\rbrace$. At this stage the reader may critically compare our
set of assumptions with the assumptions he would wish to make. At any rate, the following discussion employs simplifying assumptions and a 
more refined analysis has to take care of the neglected terms.

At the stage we arrived at with the above formulae for the backreaction term and the averaged scalar curvature, it is clear what is the potential contribution of the various terms involved. 
In particular, if, in line with the above analysis, we now assume that:\\
 \emph{(i)} 
\begin{equation}
V_{\varrho} [\CE] =  0\;;\; V_{\varrho} [\CM] = 0
\;\;,
\end{equation} 
namely that we are describing the region of near homogeneity $\CD$ in a frame comoving with matter\footnote{These, henceforth neglected, terms would be relevant in the
interpretation advanced by Wiltshire and his collaborators \cite{wiltshire05a,wiltshire05b,wiltshire07a,wiltshire07b,wiltshire07c,wiltshire07d}.};\\
\emph{(ii)}
\begin{equation}
\Omega _{GW}^{\CD}
\ll1\;,
\end{equation}
\emph{i.e.}, absence of significant gravitational radiation of cosmological
origin,
then we finally obtain for the total kinematical backreaction parameter (\ref{idealizedQ}):
\begin{eqnarray}
\fl
\Omega^\CD_\CQ \simeq 
\lambda_\CM \,
\left[8\,\left(\frac{L_{\nabla\theta,\nabla\theta}^{2}}{L_{\delta H_{\CM}}^{2}}\right)-1\right]\,\frac{\delta ^{2}H_{\CM}}{H_{\cal D }^{\,2}}+(1-\lambda_\CM)\,
\left[8\,\left(\frac{L_{\nabla\theta,\nabla\theta}^{2}}{L_{\delta H_{\CE}}^{2}}\right)-1\right]\,\frac{\delta ^{2}H_{\CE}}{H_{\cal D }^{\,2}}\nonumber\\
-\lambda_\CM (1-\lambda_\CM) \frac{(H_\CE - H_\CM )^2 }{H_\CD^2 }\;.
\end{eqnarray}
This provides a reliable estimate  in terms of the natural physical parameters involved. 
As expected, 
the shear terms (responsible for the factors proportional to  $L_{\nabla\theta,\nabla\theta}^{2} / L_{\delta H_{\CM}}^{2}$ and $L_{\nabla\theta,\nabla\theta}^{2} / L_{\delta H_{\CE}}^{2}$)
tend to attenuate the overall negative contribution of the  fluctuations of the Hubble parameter. In a self--gravitating system with long--ranged 
interactions it is difficult to argue, if the $\nabla \theta $--$\nabla \theta $ current correlation length  $L_{({\nabla\theta},\,{\nabla\theta})}$ is smaller, larger or comparable to the expansion fluctuation length $L_{\delta H_{\CD}}$.
On the one hand,  
the above estimate indicates that the attenuation mechanism due to shear fluctuations cannot compensate for the {\em global} (i.e. on the homogeneity scale) negative contribution generated by $\delta ^{2}H_{{\cal E}}/ H_{{\cal D}}^{2}$ and $\delta ^{2}H_{{\cal M}}/\,H_{{\cal D}}^{2}$, since -- due to the 
assumption of existence of a scale of homogeneity and, in addition, due to our implicit assumption of a globally almost isotropic state -- the large--scale bulk
flow must cease to display correlations and will be on some large scale subordered to the global Hubble flow pattern. This latter, however, could 
display large--scale fluctuations or not, and it
is therefore to be expected that
the ratio of the correlation lengths must be small, in conformity with our setup, only if there are significant differences between the globally averaged
{\em homogeneous state} and a {\em homogeneous solution} which, this latter, features no fluctuations.   
On the other hand, we are confident that, on the void scale, the two correlation lengths are certainly comparable, and the shear term 
may also dominate over the expansion fluctuation term in $\CQ_\CE$. This latter property was generically found in the Newtonian
analysis \cite{bks} and it can be summarized by the expectation that the kinematical backreaction term would act as a 
{\em Kinematical Dark Matter} rather than as a {\em Kinematical Dark Energy} on the scale of voids (see \cite{buchert:review} for a discussion),
whereby on the global scale the domination of the expansion fluctuation term is possible and would then argue for an interpretation as 
{\em Kinematical Dark Energy}.

It appears to us that this is a fixed point of the analysis, since in any case a non--vanishing fraction $\frac{L_{\nabla\theta,\nabla\theta}^{2}}{L_{\delta H_{\CM}}^{2}}$ and $\frac{L_{\nabla\theta,\nabla\theta}^{2}}{L_{\delta H_{\CE}}^{2}}$ is going to add a negative contribution to $\CQ_\CD$ (a positive contribution to $\Omega_\CQ^\CD$) and, therefore, neglecting this term will still allow
us to provide bounds on the expected (or for some model prior necessary) fluctuations. 
As an estimate for the large--scale asymptotics we may, along the lines of this reasoning, propose the following simple formula for a rough, but to our 
opinion robust {\em lower bound} on a {\em negative} kinematical backreaction parameter, valid on the largest scales:
\begin{equation}
\label{result_idealQ}
-\Omega^\CD_\CQ <  (1-\lambda_\CM ) \,\frac{\delta ^{2}H_\CE}{H_{{\cal D}}^{2}}
+\lambda_\CM\,\frac{\delta ^{2}H_\CM}{H_{{\cal D}}^{2}}+\lambda_\CM (1-\lambda_\CM) \frac{(H_\CE - H_\CM )^2 }{H_\CD^2 }\;\;.\nonumber\\
\end{equation}
With this formula we can bound a positive kinematical backreaction $\CQ_\CD$ from above, which provides information on the maximally
expected backreaction and, in turn, on the maximally expected magnitude of the global averaged scalar curvature, to which we turn now.

\subsection{Bound on the averaged scalar curvature in the matter--dominated Late Universe}
\label{subsection:boundsR}

Adopting the above restrictions of the general formula (\ref{resultQ}) we can write down an estimate for a global {\em upper bound} on 
the sum of a {\em positive} curvature parameter (corresponding to a negative averaged curvature) and the cosmological constant parameter, valid on the
homogeneity scale. Using (\ref{hamiltonomegaD}), (\ref{result_idealQ}), and $\Omega^\CD_m = (1-\lambda_\CM ) \Omega^\CE_m + \lambda_\CM
\Omega^\CM_m\,$, we immediately get
\begin{eqnarray}
\label{result_idealR}
\Omega_\CR^\CD + \Omega^\CD_{\Lambda} = 1 - \Omega^\CD_m - \Omega^\CD_\CQ  \\
\fl
<\;\; 1 + (1-\lambda_\CM )\left[ -\Omega^\CE_m  +\frac{\delta ^{2}H_\CE}{H_{{\cal D}}^{2}}\right]
+ \lambda_\CM \left[-\Omega^\CM_m + \frac{\delta ^{2}H_\CM}{H_{{\cal D}}^{2}}\right]+
\lambda_\CM (1-\lambda_\CM)\frac{(H_\CE - H_\CM )^2 }{H_\CD^2 } \;.\nonumber
\end{eqnarray}
The above formula demonstrates that the major players in a discussion on the maximal magnitude of the averaged scalar curvature will be 
(i) the volume fraction of occupied regions, (ii) the matter densities on the two regional scales, and (iii) the fluctuations in the
volume--averaged Hubble parameters on the two regional scales, all normalized by the global volume Hubble rate that,
this latter, can also be determined through measurement of regional parameters in view of
$H_\CD = (1-\lambda_\CM ) H_\CE + \lambda_\CM H_\CM$.
We are now going to discuss this latter result more quantitatively.

\section{Concluding discussion}
\label{section:conclusions}

First, an important remark that is relevant irrespective of whether we use a simple or a more refined estimating formula, and also irrespective of how exactly we determine the right--hand--side of, e.g. the inequality (\ref{result_idealR}) from models or observations: in any case we would obtain a non--conclusive
result for the value of the averaged scalar curvature itself due to the obvious degeneracy of the left--hand--side of  Eq.~(\ref{result_idealR}).
This latter depends on the value for the cosmological constant parameter that we have to choose {\em ad hoc}. 
For example, suppose that the right--hand--side of (\ref{result_idealR}) (representing the measured physical properties) would return 
a value $0.7$ (now and in what follows concentrating on values today), we could not discriminate between the {\em concordance model},
i.e. adopting $\Omega^{\now\CD}_{\Lambda} =0.7$, and at the same time  
$\Omega_\CR^{\now\CD} = 0$, and
a backreaction--driven cosmology with an intrinsic physical curvature parameter $\Omega_\CR^{\now\CD} = 0.7$ 
and vanishing cosmological constant. Such a value would therefore be compatible with two models of very different physical nature. 
An estimate would only be conclusive by setting a prior on the value of the cosmological constant. Only in the case where we
would ``exorcise'' a cosmological constant,  
$\Omega^{\now\CD}_{\Lambda} =0$, a measurement of the right--hand--side of (\ref{result_idealR}) would be conclusive, since we could
clearly discriminate between a zero--curvature universe model and a model with large curvature parameter. This degeneracy provides an obstacle
especially in the expected cases of (i) the need of a dominating positive cosmological constant in the concordance model, and (ii)  a substantial negative averaged curvature in a backreaction--driven cosmology (since in that case, by definition,
$\Omega_\CR^{\now\CD} > 0$). However, despite this degeneracy, the physical status of the {\em concordance model} is weaker, since a
conspiracy is required between the scale--independent value of the cosmological constant and the scale--dependent physical averages that
can in principle be measured on any scale: even if such a conspiracy would hold on the largest scale $\CD$, it is likely to be violated on smaller
scales, a remark that essentially mirrors the status of a ``fitting model'' compared to a physical model for the averaged variables.

Notwithstanding, an estimate of the right--hand--side of (\ref{result_idealR}), or the more general formulae derived in this paper, is possible and useful, and we are now suggesting strategies for 
its realization.

\subsection{Strategy 1 -- numerical simulations and analytical models}

Despite the fact that we wish to estimate Riemannian properties of the present--day Universe, the derived estimates for the large--scale averaged
scalar curvature are determined by average properties on the scales of voids and matter--dominated regions. This in turn would suggest, as a first
approximation, to determine the required {\em regional} parameters from a standard Newtonian N--body simulation. Thereby we accept to ignore the differences to the
values that would be obtained for the Riemannian averages, but we expect to get an idea for the relative magnitudes of matter and expansion fluctuations within
a well--studied framework. The determination of the regional parameters has to be done by controlled N--body simulations, this control respecting the
mutual dependence of all the parameters involved: the volume fraction of matter--dominated regions, the calculation of averages and their fluctuation
properties; both sensibly depend on spatial and particle resolution as well as on definitions of, e.g. void--finding algorithms and the employed thresholds on what we consider as over-- and underdensities. In any case it is important to determine values within a {\em single} set of priors on
the construction of the simulation and attaching a precise meaning to the involved parameters.

Alternatively, an analytical estimate using relativistic models for the inhomogeneities can be investigated, first using perturbation theory along the lines
of \cite{kolbetal,lischwarz1,lischwarz2,juliane1}, and its nonlinear extensions \cite{singh4}, or studying constraints on the size of voids from swiss cheese--type or peak models \cite{kolbetal_swisscheese1,
kolbetal_swisscheese2,rasanen:peak}. In this context we are currently generalizing
the Newtonian analysis \cite{bks}, which is built on a relativistic generalization of a non--perturbative evolution model. In this model, the (root of the) metric fluctuations are calculated perturbatively, which is sensible, since the amplitude of metric perturbations is indeed small. Combining this with the exact averages then allows to take into account that the metric derivatives may not be small. 

In all these cases we can directly estimate the shear terms, since the assumption of a frame comoving with
matter is adopted.

Related to the time--evolution properties in N--body simulations or explicit analytical models we may also approach the problem dynamically
from the point of view of effective (spatially averaged) Einstein equations. 
We may focus on the dynamical late--time properties of typical regions on the regional scales. As an example we may 
exploit the kinematically averaged Einstein equations \cite{buchert:grgdust}, valid on any scale,  
and roughly estimate their time--asymptotics for a typical matter--dominated region. The aim would consist in understanding gravitationally bound systems
and a corresponding virial equilibrium assumption that would constrain the involved energies. For example, if such a virialized state is 
characterized by a stationary ``stable--clustering'' volume $V_\CM$, 
then the averaged Raychaudhuri equation \cite{buchert:grgdust} on $\CM$ provides a relation between the
averaged matter density and the kinematical fluctuations\footnote{We have inserted a phenomenological term $\Psi_\CM$, together with the 
corresponding adimensional parameter, as an average over
kinematical terms that include other forces relevant on regions $\CM$. This term could have any sign and would be associated with velocity dispersion
(compare \cite{buchertdominguez} for a derivation of the corresponding non--averaged terms in a Newtonian setting -- note also that vorticity 
is an important stabilization term that would have to be included in such a constraint).}:
\begin{equation}
\label{virial}
4\pi G \averageM{\varrho} = \CQ_\CM +\Psi_\CM \;\;;\;\;\Omega_m^\CM = - 4 \left(\Omega_\CQ^\CM + \Omega_{\Psi}^\CM \right)\;\;.
\end{equation}
Such a condition would replace the need for estimating fluctuations on typical domains $\CM$.
With neglection of the shear term and terms contained in $\Psi_\CM$ the above relation would read $4\pi G \averageM{\varrho} \cong 6\,\delta ^{2}H_\CM$, or
$\Omega_m^\CM \cong 24\, \delta ^{2}H_\CM$, which still requires estimation of $\Omega_m^\CM$.

\subsection{Strategy 2 -- observations}

The most direct, a priori model--independent strategy is, of course, to determine the values of the required parameters through observations. 
While this strategy is in principle free of priors on the spatial geometry, the {\em interpretation} of observational results has to involve model
priors. Note that, at any rate, a model--dependence arises due to the fact that a volume--averaged value needs information on metrical properties, i.e. 
the volume depends on the geometry which is not directly observable (see, e.g., \cite{hellaby:mass,luhellaby} for strategies related to the determination
of metrical properties).
Actually all ingredients that are needed in our estimates are controversal in the literature, and accordingly an observational determination would
involve an ample range of values for the parameters and would suffer from the fact that different methods and interpretation mechanisms are 
necessarily involved and consequently would imply a loss of control on the mutual dependence of the parameters.

\subsection{Illustration}

Let us now put different model assumptions on the averaged curvature into perspective by just choosing some numbers for the
physical properties on the regional domains.
We write down the bound (\ref{result_idealR}) for (i) the {\em concordance model} with (let us take rough numbers)
$\Omega^{\now\CD}_m =0.3$, $\Omega^{\now\CD}_\CR = 0$, 
$\Omega^{\now\CD}_{\Lambda} = 0.7$, and (ii) an extreme  {\em backreaction--driven model}  with  
$\Omega^{\now\CD}_m =0.3$, $\Omega^{\now\CD}_\CR = 1$, and 
$\Omega^{\now\CD}_{\Lambda} = 0$; we obtain on the left--hand--side of (\ref{result_idealR}) the value LHS$=0.7$ for the former 
and LHS$=1$ for the latter. The right--hand--side of (\ref{result_idealR}) reads:
\begin{equation}
\label{RHS0}
\fl
{\rm RHS} =  1- 0.3 + (1-\lambda_{\now\CM}) \, \alpha_{\now\CE} + \lambda_{{\now\CM}} \alpha_{\now\CM} + \lambda_{\now\CM} (1-\lambda_{\now\CM}) \frac{(H_{\now\CE} - H_{\now\CM} )^2 }{H_{\now\CD}^2} \;\;,
\end{equation}
$$
{\rm with}\qquad\qquad\alpha_{\now\CE} :=
\frac{\delta ^{2}H_{{\cal E}}}{H_{{\cal D}}^{2}}(t_0)\;\;;\;\;\alpha_{\now\CM} :=\frac{\delta ^{2}H_{{\cal M}}}{H_{{\cal D}}^{2}}(t_0)\;\;.
$$
Here, we have first ignored the possibility of choosing different values for the regional matter--density parameters to ease the comparison. 
To make the average expansion properties concrete, we employ our scaling models for $H_{\now\CE}$ and $H_{\now\CM}$,
see Eq.~(\ref{scalingmodel1})ff., 
with a transition value $\lambda_{\now\CM}^{cr} \cong 0.1$.

The reader may now specify the free parameters involved ($\lambda_{\now\CM}$, $\alpha_{\now\CE}$, $\alpha_{\now\CM}$).
Considering a maximal range for the volume fraction, $0 < \lambda_{\now\CM} < 0.5$, we
then obtain RHS = $0.7 +  \alpha_{\now\CE}$ for a completely empty universe model, and RHS = $ 0.7 + 0.5 (  \alpha_{\now\CE}  +  
\alpha_{\now\CM} ) + 0.25 \cdot 0.09$, for a volume--equipartition of over-- and underdense regions. This shows how 
global constraints imposed by our model priors would constrain the fluctuation terms. E.g.,
a value of $\alpha_{\now\CE}= 0.3$ would be the minimally necessary fluctuation amplitude of the expansion rate on the void scale, if the
expectation from the extreme backreaction--driven model, that has practically emptied the Universe, is met.
For a non--zero $\lambda_{\now\CM}$ we have to also specify $\alpha_{\now\CM}$.

As another example let us now assume different density parameters on the different regions.
For a range of values for the volume fraction of occupied regions $\lambda_{\now\CM} = 0.2; 0.3; 0.4$ (remember that this parameter
is the relativistic volume fraction), we for example get with the priors on the global density parameter, $\Omega^{\now\CD}_m \approx 0.3$, and the density in
voids $\Omega^{\now\CE}_m \cong 0.03$ (i.e. roughly 10 percent
of the average density is found in voids) the corresponding values  $\Omega^{\now\CM}_m = 1.38; 0.93; 0.71$.
By working out a stationarity assumption in line with (\ref{virial}), we could then find $\alpha_{\now\CM}$ directly from the density parameter in 
matter--dominated regions. 

\bigskip

At this stage we leave it to the reader to
exploit the result further and to understand
the relations of assumed or measured numbers to the global model priors according to his/her experience.

\subsection{Concluding remarks on the concordance model}

In order to reconcile the standard concordance model with the estimated bounds on the averaged scalar curvature, the above discussion indicates that
this would be difficult, since for an assumed cosmological constant parameter of $0.7$, the curvature parameter acquires {\em additive}
pieces due to the fluctuations of the Hubble parameter on the two regional scales. 
Overall, the above (rough) discussion would suggest that, independent of the prior on a specific value of the cosmological constant parameter in between $0$ and a value that compensates the matter density parameter (in our example $0.7$),
the averaged curvature would be negative.

The remaining question is, how robust this latter result is, given our assumptions. For this purpose we are taking the role of advocating the 
standard concordance model by pushing the estimates to their extremes. For this end
we have to argue that all the additional contributions due to inhomogeneities are negligible, so that our formulae in fact reduce to the standard Hamiltonian constraint
for a homogeneous distribution of matter and curvature. Only in this case the concordance model, i.e. a zero--curvature universe
model, can be reconciled with the {\em physical space}. 
This statement implies that there must be a conspiracy between (A1) the shear terms on 
{\em both} regional scales, (A2) the expansion terms on {\em both} regional scales, and (A3) the difference between the Hubble rates on those 
regional scales, such that the overall contribution of these terms cancels on the homogeneity scale.
Alternatively, we could conjecture that (B1) the magnitudes of the involved fluctuation amplitudes in our formulae are quantitatively negligible
and (B2) that $H_\CE \cong H_\CM$. However, these latter options are unrealistic, since we would have to require negligible fluctuation amplitudes also on regional scales (ad B1), and that e.g. a cluster of galaxies participates in the full Hubble flow (ad B2).

Whether the above conspiracy (A) could hold is difficult to test. However, we know that it holds in a Newtonian model due to boundary conditions \cite{buchertehlers}, 
\cite{bks}, i.e. this conspiracy is suggesting that 
the present--day Universe can be effectively described in Newtonian terms \cite{wald}, i.e. it is equivalent to saying that the structure formation history had no impact on the evolution
of the averaged curvature, and fluctuations have been decoupled from the geometry (being a dynamical variable in general relativity) until today.
Although we cannot exclude this possibility, it relies on a fine--tuning assumption that can be physically justified for a Euclidean universe model,
where the curvature distribution trivially satisfies an equipartition law (since there is no curvature). 
As we have discussed, we cannot expect such an equipartition of curvature in the general case, especially when the partitioning is biased towards
a volume--dominance of underdense regions (negative curvature).

Thus, to reconcile the standard concordance model with the physical average of the present--day Universe is essentially equivalent with postulating 
an equipartition law for the scalar curvature on the homogeneity scale $\CD$, hence a fine--tuning assumption on the curvature distribution. Here, one should be aware of the fact
that such an equipartition law must be the result of a conservation law for the scalar curvature in a dynamical situation, similar as an equipartition of
the density distribution on $\CD$ being the result of restmass conservation. We know that such a conservation law does not exist in general. In the case of a dust matter model it is replaced by the condition (see \cite{buchert:review} for a detailed discussion):
\begin{equation}
\label{integrability}
\frac{1}{V_\CD^2}\partial_t \left(\,{\CQ}_\CD \,V_\CD^2 \,\right) 
\;+\; \frac{1}{V_\CD^{2/3}} \;\partial_t \left(\,\average{\CR}V_\CD^{2/3} \,
\right)\,=0\;,
\end{equation}
i.e. a particular dynamical combination of kinematical fluctuations and averaged scalar curvature is conserved, not the averaged scalar curvature itself,
that here would be represented by the conservation of the Yamabe functional $\average{\CR}V_\CD^{2/3}$. This does not mean that such a conservation law is excluded, but it would be equivalent to an
uncoupled evolution of fluctuations in a constant--curvature space section, since $\partial_t \left(\,\average{\CR}V_\CD^{2/3}\right) = 0$ implies 
$\average{\CR} \propto V_\CD^{-2/3}$. It is interesting that the mechanism of a backreaction--driven model relies on a coupled evolution between
${\CQ}_\CD$ and $\average{\CR}$ (thoroughly discussed in \cite{buchert:static,morphon}), hence, it genuinely violates an equipartition of curvature.

\subsection{Summary}

We conclude that we need a fine--tuning assumption on the scalar curvature distribution (an equipartition law on the scale of homogeneity) in order to reconcile the concordance model as a viable model for the physical properties of the present--day Universe. Dynamically,
this hypothesis implies an uncoupled evolution of kinematical fluctuations and intrinsic curvature. If such a hypothesis is not adopted,
our investigations point to an overall negative averaged scalar curvature.
Measuring fluctuations in the volume Hubble rate on the two regional scales together with the volume fraction of occupied 
regions would allow us to support or to rule out a large negative averaged scalar curvature, required for an extreme backreaction--driven model, only if
$\Lambda =0$. The crux in this consideration is the degeneracy by assuming a non--zero cosmological constant.
However, if the large value for the fluctuations needed for a 
backreaction--driven model is achievable (for calculations and discussions of estimates see \cite{bks}; \cite{rasanenrev}, Sect.3.3; \cite{singh4} and \cite{lischwarz2,lischwarz3}), there would be good reasons to shift our {\em interpretation} from the assumption of a ``curvature--compensating'' $\Lambda$ to the acceptance of a substantial negative physical curvature.

\smallskip

\subsection*{\sl Acknowledgements}

\small{\sl This work is supported 
by the Ministero dell' Universita' e della Ricerca 
Scientifica under the PRIN project `The Geometry of Integrable Systems'.
TB acknowledges support by and hospitality at the University of Pavia during working visits.
Thanks go to Martin Kerscher for valuable remarks during the preparation of the manuscript,
to Ruth Durrer, Martin Kunz and Syksy R\"as\"anen for useful discussions, and to Henk van Elst for comments
on the final manuscript.}

\end{document}